\documentclass[reprint, onecolumn]{revtex4}
\usepackage{graphicx}
\usepackage{dcolumn}
\usepackage{bm}
\usepackage{color}
\usepackage{amsmath,amsthm}
\usepackage{amssymb,bm}
\usepackage{amsbsy}
\usepackage{ulem}

\graphicspath{    {converted_graphics/}    {/}}

\begin{document}

\title{Symmetry-dependent exciton-exciton interaction and intervalley
biexciton in monolayer transition metal dichalcogenides }
\author{Hoang Ngoc Cam}
\email{hncam@iop.vast.vn}
\affiliation{Institute of Physics, Vietnam Academy of Science and Technology, 100000
Hanoi, Vietnam,}
\affiliation{Bogoliubov Laboratory of Theoretical Physics, Joint Institute for Nuclear
Research, 141980 Dubna, Moscow Region, Russia}
\author{Nguyen Thanh Phuc}
\affiliation{Department of Molecular Engineering, Graduate School of Engineering, Kyoto
University, Kyoto 615-8510, Japan}
\author{Vladimir A. Osipov}
\affiliation{Bogoliubov Laboratory of Theoretical Physics, Joint Institute for Nuclear
Research, 141980 Dubna, Moscow Region, Russia}

\begin{abstract}
The multivalley band structure of monolayer transition metal dichalcogenides
(TMDs) gives rise to intravalley and intervalley excitons. Much knowledge of
these excitons has been gained, but fundamental questions remain, such as
how to describe them all in a unified picture with their correlations, how
are those from different valleys coupled to form the intervalley biexciton?
To address the issues, we derive an exciton Hamiltonian from interpair
correlations between the constituent carriers-fermions of two excitons.
Identifying excitons by irreducible representations of their point symmetry
group, we find their pairwise interaction depending on interacting excitons'
symmetry. It is generally repulsive, except for the case excitons from
different valleys, which attract each other to form the intervalley
biexciton. We establish a semianalytical relationship between the biexciton
binding energy with exciton mass and dielectric characteristics of the
material and surroundings. Overall, by providing insight into the nature of
diverse excitons and their correlations, our theoretical model captures the
exciton interaction properties permitting an inclusive description of the
structure and energy features of the intervalley biexciton in monolayer TMDs.
\end{abstract}

\author{}
\maketitle




\section{INTRODUCTION}

Monolayer (ML) group-VI transition metal dichalcogenides (TMDs), such as MoS$%
_{2}$, MoSe$_{2}$, WS$_{2}$ and WSe$_{2}$, are two-dimensional (2D)
semiconductors with direct band gaps at the edges $K$ and $K^{\prime }$ of
the hexagonal Brillouin zone (BZ).$^{1,2}$ The reduced dielectric screening
of the Coulomb interaction$^{3}$ results in the formation of tightly bound
excitons at the $K$ and $K^{\prime }$ valleys, dominating the optical
response of the materials.$^{4,5}$ Besides the optically accessible bright
excitons, the electronic structure of ML TMDs gives rice to inaccessible
dark excitons, affecting different optical processes near the exciton
resonance.$^{6-8}$ Despite numerous works on exciton physics, a unified
picture of diverse excitons with their quantum correlations is still
lacking. Thus the understanding of such an exciton fundamental feature as
the exciton-exciton interaction remains limited. Scarce theoretical studies
consider only the intravalley interaction between identical bright excitons,
showing that it is repulsive in the exciton ground state.$^{9,10}$
Meanwhile, experiments report signatures of the intervalley biexciton in
various ML TMDs$^{11-18}$ indicating an intervalley attractive
exciton-exciton interaction. The attraction between excitons from opposite
valleys certainly has a connection with the intervalley coupling between
their constituent charge carriers via the Coulomb interaction. Several
authors groups have attempted to model the intervalley biexciton.$^{19-23}$
However, without the intervalley carrier-carrier interaction taken into
consideration, they have not succeeded. In particular, their calculations
give for the biexciton binding energy in different freestanding MLTMDs
comparable values around 20 meV, whereas experimental reports are markedly
diverse. Experiments show that the exciton-exciton interaction in ML\ TMDs
is enhanced, offering perspectives for engineering exciton-mediated optical
nonlinearities.$^{24}$ It qualitatively changes the physical picture of the
coherent light-matter interaction in the optical Stark effect.$^{25-27}$
Especially, involvement of the intervalley biexciton makes this effect
valley-dependent, giving a possibility for coherent manipulation of the
exciton valley degree of freedom in quantum information.$^{28,29}$ Thus a
comprehensive study of the exciton-exciton interaction and the intervalley
biexciton is of necessity not only for fundamentals of many-body physics but
also for promising quantum technologies applications.

To address the elusive issue, we derive an exciton Hamiltonian from
correlations between the constituent charge carriers-fermions of two
excitons, mediated by the electrostatic carrier-carrier interaction and the
Pauli exclusion principle. Identifying each exciton by an irreducible
representation of their point symmetry group, we find the exciton-exciton
interaction depending on the interacting excitons' symmetry. It is generally
repulsive, except for the case excitons from different valleys, which
attract each other. We elucidate the microscopic mechanism underlying the
intervalley exciton-exciton attraction. We ascertain a substantial
dependence of the intervalley interaction on the exciton radius, determining
the overlap degree of the wave functions of distant excitons in the momentum
space. Adopting the Kedysh potential for the carrier-carrier interaction,$%
^{3}$ we have the exciton radius as the variational parameter.$^{30,31}$ We
find it from a function established between the exciton binding energy with
its mass and the material and environment dielectric characteristics. With
values of the latter as input variables taken from experimental measures,$%
^{32,33}$ we find the intervalley interaction potential sufficiently weak to
be considered a perturbation. As a result, the estimated biexciton binding
energy has the form of an exponential function of the exciton mass and
intervalley interaction energy.$^{34}$ Its sensitivity to every input
variable can help understanding discrepancies between different experimental
measurements.$^{11-18}$ In freestanding tungsten-based MLs with light
excitons, the estimated binding energy is under 20 meV, while in ML MoSe$%
_{2} $ and MoS$_{2}$ with heavier excitons, it is about 65 meV and 53 meV,
respectively. The last number is near the appropriate experimental
measurement of Sie et al.$^{12}$ Studying the reduction of the biexciton
binding energy with increasing environment screening, we find, e.g. that the
biexciton binding energy in hBN encapsulated ML MoSe$_{2}$ falls to the
range of 24 meV, going along with recent experimental findings.$^{27}$ The
obtained dependence of the biexciton binding energy in ML TMDs on the
average dielectric constant of their immediate surroundings might serve as
guidelines for future experiments to study the biexciton feature in various
dielectric environments. On the other hand, our symmetry-dependent exciton
Hamiltonian would form a baseline for theoretical research on valley
selective nonlinear effects in a coherently driven ML TMD.

\section*{RESULTS}

\subsection*{Valley single-particle states and their interaction}

We consider an ML TMD having direct band gaps with the conduction and
valence band extrema at the $K$ and $K^{\prime }$ valleys. The point
symmetry group of the material is $D_{3h}$, but at the valleys the wave
vector group is $C_{3h}$. To exploit the group theoretical algebra
elaborated for the point group, we classify the valley Bloch states by
one-dimensional spinor (double-valued) representations of the $C_{3h}$
double group in Koster et al. notations.$^{35}$ Each spinor representation
corresponds to a definite half-integral angular momentum value including the
momentum $j$ and its projection $j_{z}$ on the $z$-axis. Thus the valley
states can be alternatively identified by those of their angular momentum $%
|\,j,j_{z}\,\rangle $, which we will refer to shortly as spin and spin
projection. Thanks to large valence band splittings,$^{36,37}$ we can
exclude the lower spin-orbit split valence band from consideration by
considering the selective excitation of the ground state A exciton. Under
this condition, we sketch the band structure of ML TMDs at the $K$ and $%
K^{\prime }$ valleys in Fig. 1, where we take the tungsten-based (WX$_{2}$)
subgroup with the order of conduction bands reverse to that in
molybdenum-based (MoX$_{2}$) subgroup, X = S, Se.$^{38}$

A resonant light field applied to a direct-gap semiconductor raises
electrons from the filled valence band into the empty conduction band. The
promotion of an electron creates a pair of electronic states, including the
conduction band electron and the empty state that it leaves in the valence
band, as schematically shown in Fig. 1. The electric dipole interaction of a
polarized light field propagating along axis $z$ with the system of created
pair states is defined by the product $\mathbf{E}_{\lambda }\mathbf{d}$ with 
$\mathbf{d}$ the system electric dipole momentum and $\mathbf{E_{\lambda }\,}
$the electric field of photons with spin projection $\lambda $ on the $z$%
-axis. The field $\mathbf{E_{\lambda }\propto \varepsilon }_{\lambda }$,
where $\mathbf{\varepsilon }_{\lambda }$ is the polarization vector
orthogonal to the direction of the light propagation. For the circularly
polarized $\sigma _{+}$ and $\sigma _{-}$ light with $\lambda =1$ and $%
\lambda =-1$ respectively, the polarization vector $\mathbf{\varepsilon }%
_{\lambda }\propto (1+\lambda i)$.$^{39}$ Thus at normal incidence, the
coupling of the optical field to the electronic system is determined by the
matrix element of the dipole momentum $d_{\pm }=(d_{x}\pm id_{y})$ between
the valence and conduction bands. Composing of components of a polar vector, 
$d_{+}$ and $d_{-}$ are transformed according to the representation $\Gamma
_{2}$ and $\Gamma _{3}$ of the $C_{3h}$ group, respectively.$^{35}$
According to the group theory selection rules$^{40}$ and multiplication
table of irreducible representations of the $C_{3h}$ double group, one has
interband matrix elements $\left\langle \Gamma _{11}|d_{+}|\Gamma
_{7}\right\rangle \neq 0$ and $\left\langle \Gamma _{12}|d_{-}|\Gamma
_{8}\right\rangle \neq 0$, and $\left\langle \Gamma _{9}|d_{+}|\Gamma
_{7}\right\rangle =\left\langle \Gamma _{10}|d_{-}|\Gamma _{8}\right\rangle
=0$. In this way, direct transitions to the $\Gamma _{11}$ ($\Gamma _{9}$)
conduction band at the $K\ $valley and to $\Gamma _{12}$ ($\Gamma _{10}$)
one at the $K^{\prime }\ $valley from the respective valence band are dipole
allowed (forbidden). The light of $\sigma _{+}$ circular polarization can
create pair states exclusively at $K$ valley, while that of $\sigma _{-}$
polarization can do this only at $K^{\prime }\ $valley. This is the
valley-dependent optical selection rule.$^{36,41,42}$ To focus on the
intervalley interaction between excitons, hereafter we will leave aside the
split-off conduction bands that are connected with the intravalley dipole
forbidden excitons (the lower bands in Fig. 1). Thus, we limit ourselves in
this paper to a simplified model with two-band schemes at the $K$ and $%
K^{\prime }$ valleys. Because excitons are Coulomb-bound electron-hole
pairs, we begin by considering the pairwise interaction among carriers. In
the second quantization representation, the system of these single-particle
states is described by Heisenberg creation and annihilation field operators $%
\Phi ^{\dagger }(\mathbf{r})$ and $\Phi (\mathbf{r})$. The operator of any
macroscopic physical quantity of the many-electron system, in particular the
number of pair states and Hamiltonian, is presented in terms of $\Phi
^{\dagger }(\mathbf{r})$\ and $\Phi (\mathbf{r})$,$^{43}$

\begin{figure}[tbp]
\centering
\includegraphics[width=4.5in,height=2.31in,keepaspectratio]{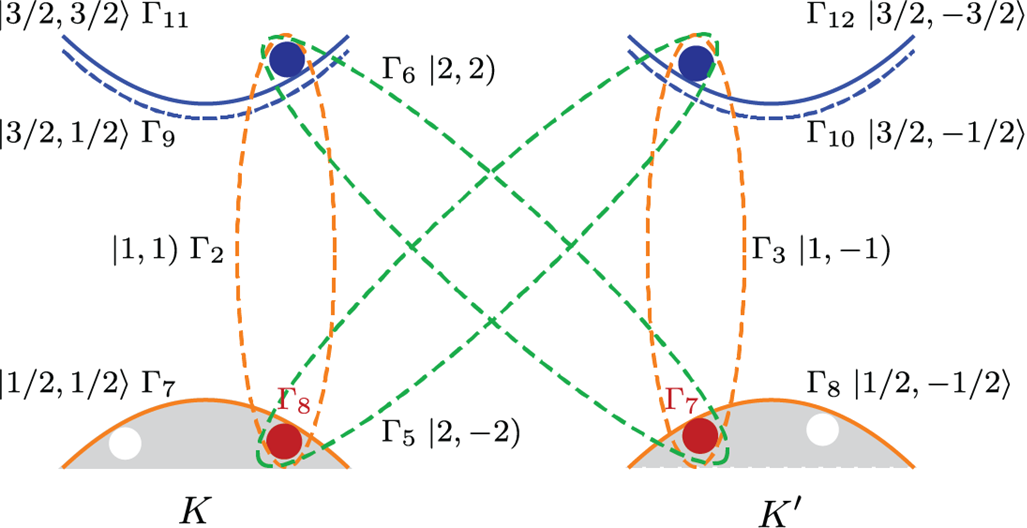}
\caption{Sketch of the band structure of ML tungsten dichalcogenides at the $%
K$ and $K^{\prime }$ valleys and related excitons. Valley states and excitons
are denoted by irreducible representations of the $C_{3h}$ double group with
the corresponding spin states shown beside. Blue, white, and dark red balls
depict conduction electrons, missing valence band electrons, and holes,
respectively. Bright and dark excitons are represented by orange and green
dashed ovals, respectively, incorporating corresponding electrons and holes.}
\label{fig:F1}
\end{figure}

\begin{equation}
N=\int d^{2}r\,\Phi ^{\dagger }(\mathbf{r})\,\Phi (\mathbf{r})
\end{equation}%
\begin{equation}
H=\int d^{2}r\,\Phi ^{\dagger }(\mathbf{r})\,\mathcal{H}_{c}\,\Phi (\mathbf{r%
})+\frac{1}{2}\int \int d^{2}r_{1}\,d^{2}r_{2}\,\Phi ^{\dagger }(\mathbf{r}%
_{1})\Phi ^{\dagger }(\mathbf{r}_{2})\,V_{K}(|\mathbf{r}_{1}-\mathbf{r}%
_{2}|)\,\Phi (\mathbf{r}_{2})\Phi (\mathbf{r}_{1})
\end{equation}%
where $\mathcal{H}_{c}$ is the Hamiltonian of a single crystal electron in
the periodic lattice potential and $V(r)$ --{\ nonlocally screened Coulomb
potential describing the symmetric pairwise interaction between in-plain
carriers. We adopt the Keldysh potential,}$^{3}${\ which can be presented in
the form \ }%
\begin{equation}
V_{K}(r)\simeq \frac{e^{2}}{\varepsilon _{0}r_{0}}\frac{\pi }{2}\left[
H_{0}\left( \frac{\kappa r}{r_{0}}\right) -Y_{0}\left( \frac{\kappa r}{r_{0}}%
\right) \right]
\end{equation}%
where $e$ is the electron charge, $\varepsilon _{0}$ -- the vacuum
permittivity, $r_{0}$ -- the effective screening length characterizing
dielectric properties of an ML TMD, $\kappa =\left( \varepsilon
_{t}+\varepsilon _{b}\right) /2$ with $\varepsilon _{t}$ and $\varepsilon
_{b}$ the dielectric constants of the encapsulating materials above and
below the ML, respectively, and $H_{0}$ and $Y_{0}$ -- Struve and Bessel
functions of the second kind. The screening length of an ML having width $d$
and dielectric constant $\varepsilon $ is defined as $r_{0}=d\varepsilon
/\left( \varepsilon _{t}+\varepsilon _{b}\right) $ with $\varepsilon
_{t}=\varepsilon _{b}=1$ (vacuum). In the strictly 2D limit $r_{0}=2\pi \chi
_{2D}$, where $\chi _{2D}$ is the 2D polarizability of the planar material.$%
^{44}$ An inspection of Eq. (3) shows that the carrier-carrier interaction
in ML\ TMDs weakens with increasing screening, either the ML screening ($%
r_{0}$) or that from the environment ($\kappa $). Thus the interaction can
be 'tuned' by selecting different immediate surroundings for the ML.

We expand field operators into the complete set of orthonormal Bloch
functions -- the eigenstates of $\mathcal{H}_{c}$,$^{45}$ $\phi _{\Gamma ,%
\mathbf{k}}(\mathbf{r})=u_{\Gamma ,\mathbf{k}}(\mathbf{r})\exp [i\mathbf{kr}%
]/\sqrt{S}$ with $S$ the sample area and $\Gamma $ the band states symmetry.
With the assumption that under resonant excitation crystal electrons are
accummulated primarily near $K$ and\ $K^{\prime }$ valleys, we can limit the
expansion to the wave vectors around the valleys,\qquad\ 
\begin{equation}
\Phi (\mathbf{r})\simeq \sum\limits_{\substack{ \Gamma =\Gamma
_{11},\,\Gamma _{7}  \\ \mathbf{p}}}{\large {e}_{\Gamma ,\,\mathbf{p}}\,}%
\phi _{\Gamma ,\,\mathbf{p}}(\mathbf{r})+\sum\limits_{\substack{ \Gamma
=\Gamma _{12},\,\Gamma _{8}  \\ \mathbf{p}^{\prime }}}{\large {e}_{\Gamma ,\,%
\mathbf{p}^{\prime }}\,}\phi _{\Gamma ,\,\mathbf{p}^{\prime }}(\mathbf{r})
\end{equation}%
where the sums are running over $\mathbf{p=k-K}$ and $\mathbf{p}^{\prime }%
\mathbf{=k-K}^{\prime }$, $p,p^{\prime }\ll \left\vert \mathbf{K}\right\vert
,\left\vert \mathbf{K}^{\prime }\right\vert $, with $\mathbf{K}$ and $%
\mathbf{K}^{\prime }$ the positions of the BZ corner points in the \textbf{k}%
-space,%
\begin{equation}
\mathbf{K=}\frac{2\pi }{3a}\left( 1,\frac{1}{\sqrt{3}}\right) ,\qquad 
\mathbf{K}^{\prime }\mathbf{=}\frac{2\pi }{3a}\left( 1,-\frac{1}{\sqrt{3}}%
\right)
\end{equation}%
($a$ is the lattice constant).$^{46}$ Putting $\hbar =1$, we call vectors $%
\mathbf{p}$ and $\mathbf{p}^{\prime }$ and also any their linear combination
the valley momenta of quasiparticles distinguising them from their crystal
momenta. In Eq. (4) ${\large e}_{\Gamma ,\mathbf{p}}$ is the annihilation
operator for an electronic state with symmetry $\Gamma $ and valley momentum 
$\mathbf{p}$ obeing fermionic anticommutation relations. By inserting Eq.
(4) to Eqs (1) and (2) we obtain the number operator and Hamiltonian in
terms of creation and annihilation operators of electronic states. It is
conventional to describe an empty electron state in valence bands as a hole
related to the state by the time-reversal transformation. The hole charge is 
$-e$, its wave vector is opposite to that of the missing valence band
electron, and its symmetry notation is complex conjugate to that of the
last. In this way, the hole going in pair with a conduction electron at the $%
K$ valley has wave vector $\mathbf{k}_{h}=\mathbf{-K-p}$ and notated by $%
\Gamma _{7}^{\ast }=\Gamma _{8}$. In Fig. 1 and others, we mark the hole
symmetry notations by dark red color. Further, as $\mathbf{p}$ and $\mathbf{p%
}^{\prime }$ are running vector-index over valence bands, which are assumed
isotropic, we can write the hole's wave vector in the form $\mathbf{k}_{h}=%
\mathbf{-K+p}_{h}$ at $K$ and $\mathbf{k}_{h}=\mathbf{-K}^{\prime }\mathbf{+p%
}_{h}^{\prime }$ at $K^{\prime }$.

As a result, we obtain the Hamiltonian of the system of electron-hole pairs
in the form

\begin{eqnarray}
H &\rightarrow &\mathcal{\mathcal{H}}_{eh}=\sum\limits_{\mathbf{p}}\left[
E_{e}(p){\large e}_{\mathbf{\Gamma }_{11}\mathbf{,\,p}}^{+}{\large e}_{%
\mathbf{\Gamma }_{11}\mathbf{,\,p}}+E_{h}(p){\large h}_{\mathbf{\Gamma }_{8}%
\mathbf{,\,p}}^{+}{\large h}_{\mathbf{\Gamma }_{8}\mathbf{,\,p}}\right] 
\notag \\
&&+\sum\limits_{\mathbf{p}^{\prime }}\left[ E_{e}(p^{\prime }){\large e}_{%
\mathbf{\Gamma }_{12}\mathbf{,\,p}^{\prime }}^{+}{\large e}_{\mathbf{\Gamma }%
_{12}\mathbf{,\,p}^{\prime }}+E_{h}(p){\large h}_{\mathbf{\Gamma }_{7}%
\mathbf{,\,p}^{\prime }}^{+}{\large h}_{\mathbf{\Gamma }_{7}\mathbf{,\,p}%
^{\prime }}\right]  \notag \\
&&+\frac{1}{2}\sum\limits_{\mathbf{q\neq 0}}V_{q}\left\{ \sum\limits_{%
\mathbf{p}_{1}\mathbf{,\,p}_{2}}\left[ {\large e}_{\mathbf{\Gamma }_{11},\,%
\mathbf{p}_{1}+\mathbf{q}}^{+}{\large e}_{\mathbf{\Gamma }_{11},\,\mathbf{p}%
_{2}-\mathbf{q}}^{+}{\large e}_{\mathbf{\Gamma }_{11},\,\mathbf{p}_{2}}%
{\large e}_{\mathbf{\Gamma }_{11},\,\mathbf{p}_{1}}\right. \right.  \notag \\
&&\left. +{\large h}_{\mathbf{\Gamma }_{8}\mathbf{,\,p}_{1}+\mathbf{q}}^{+}%
{\large h}_{\mathbf{\Gamma }_{8}\mathbf{,\,p}_{2}-\mathbf{q}}^{+}{\large h}_{%
\mathbf{\Gamma }_{8}\mathbf{,\,p}_{2}}{\large h}_{\mathbf{\Gamma }_{8}%
\mathbf{,\,p}_{1}}-2\,{\large e}_{\mathbf{\Gamma }_{11},\,\mathbf{p}_{1}+%
\mathbf{q}}^{+}{\large h}_{\mathbf{\Gamma }_{8}\mathbf{,\,p}_{2}-\mathbf{q}%
}^{+}{\large h}_{\mathbf{\Gamma }_{8},\,\mathbf{p}_{2}}{\large e}_{\mathbf{%
\Gamma }_{11}\mathbf{,\,p}_{1}}\right]  \notag \\
&&+\sum\limits_{\mathbf{p}_{1}^{\prime }\mathbf{,\,p}_{2}^{\prime }}\left[ 
{\large e}_{\mathbf{\Gamma }_{12},\,\mathbf{p}_{1}^{\prime }+\mathbf{q}}^{+}%
{\large e}_{\mathbf{\Gamma }_{12},\,\mathbf{p}_{2}^{\prime }-\mathbf{q}}^{+}%
{\large e}_{\mathbf{\Gamma }_{12},\,\mathbf{p}_{2}^{\prime }}{\large e}_{%
\mathbf{\Gamma }_{12},\,\mathbf{p}_{1}^{\prime }}+{\large h}_{\mathbf{\Gamma 
}_{7}\mathbf{,\,p}_{1}^{\prime }+\mathbf{q}}^{+}{\large h}_{\mathbf{\Gamma }%
_{7}\mathbf{,\,p}_{2}^{\prime }-\mathbf{q}}^{+}{\large h}_{\mathbf{\Gamma }%
_{7}\mathbf{,\,p}_{2}^{\prime }}{\large h}_{\mathbf{\Gamma }_{7}\mathbf{,\,p}%
_{1}^{\prime }}\right.  \notag \\
&&\left. -2\,{\large e}_{\mathbf{\Gamma }_{12},\,\mathbf{p}_{1}^{\prime }+%
\mathbf{q}}^{+}{\large h}_{\mathbf{\Gamma }_{7}\mathbf{,\,p}_{2}^{\prime }-%
\mathbf{q}}^{+}{\large h}_{\mathbf{\Gamma }_{7},\,\mathbf{p}_{2}^{\prime }}%
{\large e}_{\mathbf{\Gamma }_{12}\mathbf{,\,p}_{1}^{\prime }}\right]  \notag
\\
&&+2\sum\limits_{\mathbf{p,\,p}^{\prime }}\left[ {\large e}_{\mathbf{\Gamma }%
_{11},\,\mathbf{p}+\mathbf{q}}^{+}{\large e}_{\mathbf{\Gamma }_{12},\,%
\mathbf{p}^{\prime }-\mathbf{q}}^{+}{\large e}_{\mathbf{\Gamma }_{12},\,%
\mathbf{p}^{\prime }}{\large e}_{\mathbf{\Gamma }_{11},\,\mathbf{p}}+{\large %
h}_{\mathbf{\Gamma }_{8}\mathbf{,\,p}+\mathbf{q}}^{+}{\large h}_{\mathbf{%
\Gamma }_{7}\mathbf{,\,p}^{\prime }-\mathbf{q}}^{+}{\large h}_{\mathbf{%
\Gamma }_{7}\mathbf{,\,p}^{\prime }}{\large h}_{\mathbf{\Gamma }_{8}\mathbf{%
,\,p}}\right.  \notag \\
&&\left. \left. -\left( {\large e}_{\mathbf{\Gamma }_{11},\,\mathbf{p}+%
\mathbf{q}}^{+}{\large h}_{\mathbf{\Gamma }_{7}\mathbf{,\,p}^{\prime }-%
\mathbf{q}}^{+}{\large h}_{\mathbf{\Gamma }_{7},\,\mathbf{p}^{\prime }}%
{\large e}_{\mathbf{\Gamma }_{11}\mathbf{,\,p}}+{\large e}_{\mathbf{\Gamma }%
_{12},\,\mathbf{p}^{\prime }+\mathbf{q}}^{+}{\large h}_{\mathbf{\Gamma }_{8}%
\mathbf{,\,p}-\mathbf{q}}^{+}{\large h}_{\mathbf{\Gamma }_{8}\mathbf{,\,p}}%
{\large e}_{\mathbf{\Gamma }_{12},\,\mathbf{p}^{\prime }}\right) \right]
+h.c.\right\} ,
\end{eqnarray}%
where $V_{q}$ is the Fourier {transform} of the Keldysh potential, and $%
E_{e}(p)$ and $E_{h}(p)$ -- the single-particle electron and hole energies,
which are renormalized due to the interaction with the valence band
electrons. In the effective mass approximation and with {the energy zero
chosen at the top of the valence bands,} $E_{e}(p)=E_{g}+p^{2}/2\mu _{e}$ ($%
E_{g}$ is the band gap) and $E_{h}(p)=p^{2}/2\mu _{h}${, }where $\mu ${$_{e}$
($\mu _{h}$) is the electron (hole) effective mass}. To arrive at Eq. (6),
the Wannier simplifying assumptions justified for pair states with small
relative momenta and large space extent$^{47}$ have been used, along with
the orthonormal properties of the Bloch functions and periodicity of its
amplitudes. Yet, we restrict ourselves to the pairwise interaction
processes, which conserve the number of electron-hole pairs. As expected,
the electrostatic interaction among valley carriers includes their
intravalley and intervalley interactions.

\subsection*{Diverse excitons and their pairwise interaction}

Assuming that the main contribution to excitons is from the band states in
the vicinity of the $K$\ and $K^{\prime }$ points, we have four symmetry
types of the ground state exciton in the model under consideration according
to the multiplication table of irreducible representations of the $C_{3h}$
double group.$^{35}$ That is two intravalley excitons, $\Gamma _{2}=\Gamma
_{11}\otimes \Gamma _{8}$\ at the $K$ valley and $\Gamma _{3}=\Gamma
_{12}\otimes \Gamma _{7}$\ at the $K^{\prime }$ valley, and two intervalley
excitons, $\Gamma _{6}=\Gamma _{11}\otimes \Gamma _{7}$\ and $\Gamma
_{5}=\Gamma _{12}\otimes \Gamma _{8}$, depicted respectively by orange and
green dashed ovals in Fig. 1. An exciton with symmetry $\Gamma _{x}=\Gamma
_{e}\otimes \Gamma _{h}$ and center-of-mass\ (total) valley momentum $%
\mathbf{\mathcal{K}}$ is defined as a superposition of the pair states
having the same total valley momentum with electrons and holes from the band
of symmetry $\Gamma _{e}\ $and $\Gamma _{h}$, respectively. From the
relationship between basis functions of relevant irreducible
representations, we have the relation between exciton symmetry states and
those of corresponding electron-hole pairs, 
\begin{eqnarray}
A_{\Gamma _{2},\mathbf{\mathcal{K}}}^{+}\,|\,\mathbf{0\,)} &=&\frac{1}{\sqrt{%
S}}\sum_{\mathbf{p}_{e},\,\mathbf{p}_{h}}{\large \delta }(\mathbf{p}_{e}+%
\mathbf{p}_{h},\mathbf{\mathcal{K}})\,\digamma (\alpha \mathbf{p}_{h}-\beta 
\mathbf{p}_{e})\,{\large {e}_{\Gamma _{11},\mathbf{p}_{e}}^{+}{h}_{\Gamma
_{8},\mathbf{p}_{h}}^{+}}\,|\,0\,\rangle \,,  \notag \\
A_{\Gamma _{3},\mathbf{\mathcal{K}}}^{+}|\,\mathbf{0\,)}\, &=&\frac{1}{\sqrt{%
S}}\sum_{\mathbf{p}_{e}^{\prime },\,\mathbf{p}_{h}^{\prime }}{\large \delta }%
(\mathbf{p}_{e}^{\prime }+\mathbf{p}_{h}^{\prime },\mathbf{\mathcal{K}}%
)\,\digamma (\alpha \mathbf{p}_{e}^{\prime }-\beta \mathbf{p}_{h}^{\prime
})\,{\large {e}_{\Gamma _{12},\mathbf{p}_{e}^{\prime }}^{+}{h}_{\Gamma _{7},%
\mathbf{p}_{h}^{\prime }}^{+}}\,|\,0\,\rangle ,  \notag \\
A_{\Gamma _{6},\mathbf{\mathcal{K}}}^{+}\,|\,\mathbf{0\,)} &=&\frac{1}{\sqrt{%
S}}\sum_{\mathbf{p},\,\mathbf{p}^{\prime }}{\large \delta }(\mathbf{p}+%
\mathbf{p}^{\prime },\mathbf{\mathcal{K}})\,\digamma (\alpha \mathbf{p}%
^{\prime }-\beta \mathbf{p})\,{\large {e}_{\Gamma _{11},\mathbf{p}}^{+}{h}%
_{\Gamma _{7},\mathbf{p}^{\prime }}^{+}}\,|\,0\,\rangle ,  \notag \\
A_{\Gamma _{5},\mathbf{\mathcal{K}}}^{+}|\,\mathbf{0\,)}\, &=&-\frac{1}{%
\sqrt{S}}\sum_{\mathbf{p},\,\mathbf{p}^{\prime }}{\large \delta }(\mathbf{p}%
^{\prime }+\mathbf{p},\mathbf{\mathcal{K}})\,\digamma (\alpha \mathbf{p}%
-\beta \mathbf{p}^{\prime })\,{\large {e}_{\Gamma _{12},\mathbf{p}^{\prime
}}^{+}{h}_{\Gamma _{8},\mathbf{p}}^{+}}\,|\,0\,\rangle .
\end{eqnarray}%
Here $|\,0\,\rangle $ denotes the semiconductor ground state in the
electron-hole presentation with the valence bands filled and the conduction
bands empty, $|\,\mathbf{0\,)}$ -- that of the exciton space, $A_{\Gamma ,%
\mathbf{\mathcal{K}}}^{+}$ -- the creation operator for the exciton with
symmetry $\Gamma $ and valley total momentum $\mathbf{\mathcal{K}}$, $%
\digamma (\mathbf{p})$ -- the momentum space wave function of the
electron-hole relative motion in the ground state exciton, and $\alpha =\mu
_{e}/\mu _{x}$ ($\beta =\mu _{h}/\mu _{x}$) -- the electron-to-exciton
(hole-to-exciton) mass ratio ($\mu _{x}=\mu _{e}+\mu _{h}$). We see that the
relation of the exciton valley total momentum, $\mathbf{\mathcal{K=}}$ $%
\mathbf{p}_{e}+\mathbf{p}_{h}$, and relative one, $\alpha \mathbf{p}%
_{h}-\beta \mathbf{p}_{e}$, to their crystal counterparts depends on the
symmetry type. The valley relative momentum differs from its crystal
counterpart by vector $\mathbf{K}$, $\mathbf{K}^{\prime }$, $\alpha \mathbf{K%
}+\beta \mathbf{K}^{\prime }$ and $\alpha \mathbf{K}^{\prime }+\beta \mathbf{%
K}$ for the symmetry $\Gamma _{2}$, $\Gamma _{3}$, $\Gamma _{5}$ and $\Gamma
_{6}$, respectively. Regarding the total momentum, the valley and crystal
counterparts are the same for the intravalley excitons, while for the
intervalley $\Gamma _{5}$ and $\Gamma _{6}$\ ones they differ from each
other by vector $\mathbf{K-K}^{\prime }$and $\mathbf{K}^{\prime }-\mathbf{K}$%
, respectively. With such large crystal momenta, the intervalley excitons
cannot be optically accessible, referred to as momentum-dark excitons. By
symmetry, they are not dipole allowed either. In fact, in line with the
group theory selection rules, under excitation by the $\sigma _{+}$ ($\sigma
_{-}$) light, the transition from the ground state $|\,\mathbf{0\,)}$
(described by the unit representation) is possible only to the exciton state
with the symmetry as that of the dipole momentum $d_{+}$ ($d_{-}$). Thus,\
under the $\sigma _{+}$ light, only the $\Gamma _{2}$ exciton at the $K$
valley is dipole active (bright), while under the $\sigma _{-}$ light such
is the $\Gamma _{3}$ exciton at the $K^{\prime }$ valley. In this way, the
symmetry notation of a bright exciton incorporates both its spin and valley
index: the $\Gamma _{2}$ ($\Gamma _{3}$) exciton is the $K$ ($K^{\prime }$)
valley exciton with spin projection $1$ ($-1$) as that of the photon with
whom it interacts. We note that excitons, consisting of two half-integral
spin carriers, are characterized by single-valued representations of the $%
C_{3h}$ group corresponding to integral spin states $|\,J,J_{z}\mathbf{\,)}$%
. Thus, the bright $\Gamma _{2}$ and $\Gamma _{3}$ (dark $\Gamma _{6}$ and $%
\Gamma _{5}$) excitons are the spin $1$\ ($2$) ones with spin projection $1$%
\ and $-1$\ ($2$\ and $-2$), respectively (see Fig. 1). The heavy hole
exciton in III-V quantum wells has the same spin states.$^{48,49}$ In
conventional 2D and 3D direct gap semiconductors with two simple (only
twofold spin-degenerate) bands, there are four states of spin 1 and spin 0
excitons.$^{50-52}$ The last can be well separated in energy, e.g. in bulk Cu%
$_{2}$O, then they are called ortho- and paraexciton, respectively. With the
dipole allowed, or quadrupole allowed in Cu$_{2}$O, interband transition,
states $|\,1,1\mathbf{\,)}$ and $|\,1,-1\mathbf{\,)}$ are bright and $|\,1,0%
\mathbf{\,)}$ and $|\,0,0\mathbf{\,)}$ are dark. The difference is, in
conventional semiconductors, excitons are all intravalley with comparable
crystal momenta near the BZ center. While keeping symmetry notations for
consistency, we will refer to excitons in ML TMDs by their spin as the need
arises for drawing an analogy with their traditional counterparts.

It is straightforward to obtain the reverse to relations of Eq. (7) by the
use of the orthonormalization and completeness of the system of the exciton
envelope functions. In the case of selective excitation of the lowest
exciton under consideration, to each pair state there corresponds an exciton
in the following way,%
\begin{eqnarray}
{\large {e}_{\Gamma _{11},\,\mathbf{p}_{1}}^{+}\,{h}_{\Gamma _{8},\,\mathbf{p%
}_{2}}^{+}}|\,0\,\rangle &=&\frac{1}{\sqrt{S}}\digamma (\alpha \mathbf{p}%
_{2}-\beta \mathbf{p}_{1})A_{\Gamma _{2},\,\mathbf{p}_{1}+\mathbf{p}%
_{2}}^{+}|\,0\,),  \notag \\
{\large {e}_{\Gamma _{12},\,\mathbf{p}_{1}^{\prime }}^{+}\,{h}_{\Gamma
_{7},\,\mathbf{p}_{2}^{\prime }}^{+}}|\,0\,\rangle &=&\frac{1}{\sqrt{S}}%
\digamma \left( \alpha \mathbf{p}_{2}^{\prime }-\beta \mathbf{p}_{1}^{\prime
}\right) \,A_{\Gamma _{3},\,\mathbf{p}_{1}^{\prime }+\mathbf{p}_{2}^{\prime
}}^{+}|\,0\,),  \notag \\
{\large {e}_{\Gamma _{11},\,\mathbf{p}}^{+}\,{h}_{\Gamma _{7},\,\mathbf{p}%
^{\prime }}^{+}}|\,0\,\rangle &=&\frac{1}{\sqrt{S}}\digamma \left( \alpha 
\mathbf{p}^{\prime }-\beta \mathbf{p}\right) \,\,A_{\Gamma _{6},\,\mathbf{p}+%
\mathbf{p}^{\prime }}^{+}|\,0\,),  \notag \\
{\large {e}_{\Gamma _{12},\,\mathbf{p}^{\prime }}^{+}\,{h}_{\Gamma _{8},\,%
\mathbf{p}}^{+}}|\,0\,\rangle &=&-\frac{1}{\sqrt{S}}\digamma \left( \alpha 
\mathbf{p}-\beta \mathbf{p}^{\prime }\right) \,A_{\Gamma _{5},\,\mathbf{p}+%
\mathbf{p}^{\prime }}^{+}|\,0\,).
\end{eqnarray}

Let us consider an ML TMD excited at the lowest exciton energy by an
ultrashort $\sigma _{+}$ circularly polarized laser pulse. The pulse
excitation generates coherent superpositions of electron-hole pairs
corresponding to the $\Gamma _{2}$ exciton with their population described
by function $\left\vert \digamma \left( \mathbf{p}\right) \right\vert ^{2}$
[see Eq. (7)]. The excitation is assumed to be sufficiently weak that pairs
density $n$ remains low, $na_{x}^{2}\ll 1$ ($a_{x}$ is the exciton radius).
As extended quasiparticles, the carriers created predominantly in the $K$
valley undergo rapid scattering among themselves via the carrier-carrier
interaction spreading in the $\mathbf{k}$ space.$^{53}$ In a time of tens to
hundreds of femtoseconds, which is typical for the carrier-carrier
scattering,$^{54}$ carrier populations in nonequivalent valleys might be
equalized. In parallel with the carriers' pairwise scattering, the exciton
formation takes place. Strong Coulomb correlations among carriers due to
reduced screening leading to stable excitons in ML\ TMDs must also result
in\ their mutual interaction already at low density. Strictly speaking, the
exciton-exciton interaction arises when there are two electron-hole pairs in
the system because of interpair correlations. These correlations produce two
qualitative changes to excitons system, considered noninteracting bosons in
the linear approximation. First, one can check by the use of Eq. (7) that
they give rise to a non-bosonic correction of order $na_{x}^{2}$ to the
commutator of exciton operators. Secondly, they produce an effective
exciton-exciton interaction of the order $E_{b}na_{x}^{2}$, $E_{b}$ is the
exciton binding energy.$^{9}$ In the first nonlinear approximation in $%
na_{x}^{2}$ relevant to the low-density limit under consideration, one can
still treat excitons as bosons interacting via the effective two-body
interaction.$^{48,52}$

To formulate an exciton Hamiltonian, we start from the fermionic
electron-hole Hamiltonian $\mathcal{H}_{eh}$. We adopt here the method of
Haug and Schmitt-Rink that consists in a low-density expansion of the
electron and hole density and pair operators into products of exciton
operators.$^{55}$ We begin with the linear approximation, which is justified
for infinitesimally small $na_{x}^{2}$. Accurately, excitons are ideal
quasiparticles only in the hypothetical case when there is one electron-hole
pair in the system,%
\begin{equation}
N=\sum\limits_{\mathbf{p}}\left( {\large e}_{\mathbf{\Gamma }_{11}\mathbf{%
,\,p}}^{+}{\large e}_{\mathbf{\Gamma }_{11}\mathbf{,\,p}}+{\large h}_{%
\mathbf{\Gamma }_{8}\mathbf{,\,p}}^{+}{\large h}_{\mathbf{\Gamma }_{8}%
\mathbf{,\,p}}\right) +\sum\limits_{\mathbf{p}^{\prime }}\left( {\large e}_{%
\mathbf{\Gamma }_{12}\mathbf{,\,p}^{\prime }}^{+}{\large e}_{\mathbf{\Gamma }%
_{12}\mathbf{,\,p}^{\prime }}+{\large h}_{\mathbf{\Gamma }_{7}\mathbf{,\,p}%
^{\prime }}^{+}{\large h}_{\mathbf{\Gamma }_{7}\mathbf{,\,p}^{\prime
}}\right) =1.
\end{equation}

Under such a condition, $\mathcal{H}_{eh}$ is reduced to a Hamiltonian
obtained from Eq. (6) by dropping the terms presenting the electron-electron
and hole-hole interactions, which can take place only when $N\geq 2$. By
inserting the unit operator from Eq. (9) into the kinetic energy terms and
then applying Eqs. (8) and their hermitic conjugates to the obtained product
of four operators and also to the electron-hole interaction terms, we recast
Hamiltonian $\mathcal{H}_{eh}$\ of the electron-hole system in the linear
approximation into the exciton representation, $\mathcal{H}%
_{0}=\sum\limits_{\Gamma ,\mathbf{\mathcal{K}}}E_{x}(\Gamma ,\mathbf{%
\mathcal{K}})A_{\Gamma ,\,\mathbf{\mathcal{K}}}^{+}\,A_{\Gamma ,\,\mathbf{%
\mathcal{K}}}$. Here and in the following the sum variable $\Gamma $ runs
over four exciton symmetry states $\Gamma _{2},\Gamma _{3},\Gamma
_{5},\Gamma _{6}$, unless noted otherwise. To obtain $\mathcal{H}_{0}$ in
the form of the sum of energies of four excitons, we have passed from the
electron and hole momenta to the exciton center-of-mass (total) and relative
momenta, then used the completeness property of the system of the exciton
envelope functions. The symmetry dependence of the exciton energy $%
E_{x}(\Gamma ,\mathbf{\mathcal{K}})=E_{g}-E_{b}+\mathbf{\mathcal{K}}%
^{2}/2\mu _{x}$ is connected with that of $\mathbf{\mathcal{K}}$, defining
the kinetic energy of the exciton center-of-mass free motion. Meanwhile,
energy $-E_{b}$ of the internal relative electron-hole motion in the
exciton, which is the solution of the effective mass approximation equation,%
\begin{equation}
\frac{\mathbf{p}^{2}}{2\mu _{r}}\digamma (\mathbf{p})-\sum\limits_{\mathbf{%
q\neq 0}}\,V_{\mathbf{q}}\,\digamma (\mathbf{p}-\mathbf{q})=-E_{b}\digamma (%
\mathbf{p})
\end{equation}%
is the same for all exciton symmetry types ($\mu _{r}$ is the exciton
reduced mass). The modification of the Coulomb interaction to the Keldysh
form results in the deviation of the exciton spectrum from the usual
hydrogenic one.$^{30,31}$ Presenting the real space variational exciton
envelope function in the conventional form $f(r)=\left[ 2/(\pi a_{x}^{2})%
\right] ^{1/2}\exp [-r/a_{x}]$, we obtain the kinetic and interaction
energies of the relative electron-hole motion in the exciton, and with them
the exciton binding energy, in an analytical form,

\begin{eqnarray}
E_{b}(a_{x}) &=&-\frac{1}{2\mu _{r}a_{x}^{2}}+\frac{e^{2}}{\varepsilon
_{0}r_{0}}\frac{2r_{0}}{\kappa a_{x}}\left\{ \frac{\left( 1-2r_{0}/\kappa
a_{x}\right) }{\left[ 1+\left( 2r_{0}/\kappa a_{x}\right) ^{2}\right] }%
\right.  \notag \\
&&\left. +\frac{\left( 2r_{0}/\kappa a_{x}\right) ^{2}}{\left[ 1+\left(
2r_{0}/\kappa a_{x}\right) ^{2}\right] ^{3/2}}\ln \left[ \left( 1+\frac{%
\kappa a_{x}}{2r_{0}}\right) \left( 1+\sqrt{1+\left( \frac{2r_{0}}{\kappa
a_{x}}\right) ^{2}}\right) +\frac{2r_{0}}{\kappa a_{x}}\right] \right\} .
\end{eqnarray}

\begin{figure}[tbp]
\centering
\includegraphics[width=6.9in,height=2.49in,keepaspectratio]{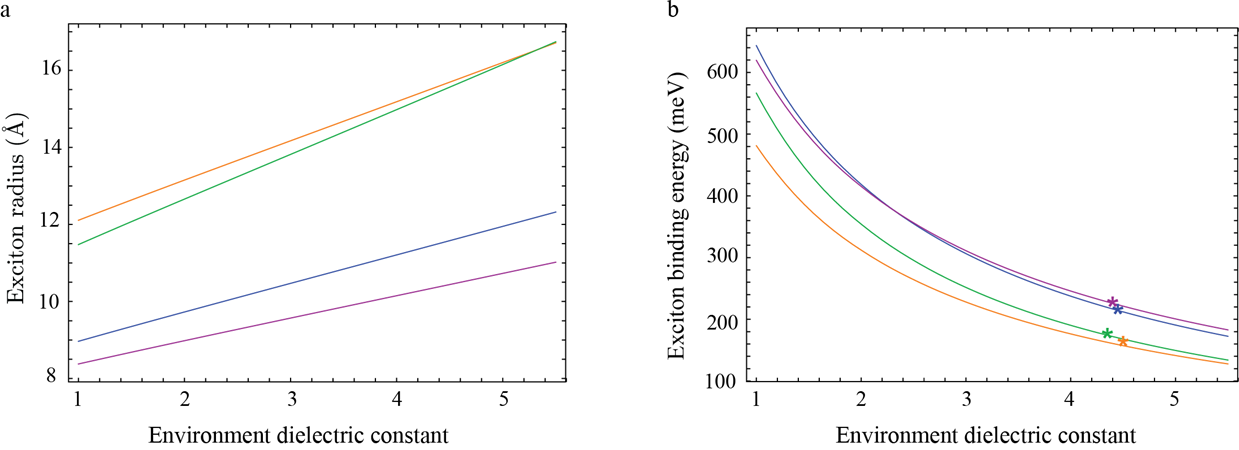}
\caption{Variation of the exciton radius (a) and binding energy (b) in ML
TMDs with the environment average dielectric constant. The orange lines are
for ML WSe$_{2}$ ($\protect\mu _{r}=0.2m_{0},r_{0}=45$ \AA ), green lines --
for ML WS$_{2}$ ($\protect\mu _{r}=0.175m_{0},r_{0}=34$ \AA ), blue lines --
for ML MoS$_{2}$ ($\protect\mu _{r}=0.275m_{0},r_{0}=34$ \AA ), and purple
lines -- for ML MoSe$_{2}$ ($\protect\mu _{r}=0.35m_{0},r_{0}=39$ \AA ). The
small stars with the respective colors depict the corresponding experimental
amounts for $E_{b}$.$^{33}$}
\label{fig:F2}
\end{figure}

From here, one can find the exciton radius as the variational parameter and
the corresponding exciton binding energy for any set of input variables,
including exciton mass $\mu _{r}$ and dielectric characteristics $r_{0}$ and 
$\kappa $. Unlike $\mu _{r}$ and $r_{0}$ as inherent features of an ML TMD,
the environment average dielectric constant can be tuned by changing
encapsulating materials above and below the ML. Thus Eq. (11) provides the
possibility to adjust exciton binding energy and space extent, which is an
advantage of ML TMDs compared to conventional 2D semiconductors. In the
last, with the regular Coulomb carrier-carrier interaction, the exciton
binding energy and radius are related to each other as $E_{b}a_{x}=e^{2}/%
\varepsilon _{0}\varepsilon $,$^{56}$ and the kinetic and electron-hole
interaction energies equal $E_{b}$ and $-2E_{b}$, respectively, independent
of the material. With the Keldysh potential, the electron-hole interaction
energy can be regulated by tuning the potential strength, ceasing to be
commensurate with $E_{b}$. The ratio between its absolute quantity and the
kinetic energy considerably increases, depending on the mass and dielectric
screening parameters. For freestanding ML TMDs, the ratio varies in the
interval 4 -- 5 in different ML TMDs, indicating the exciton robustness due
to reduced screening. Naturally, excitons are stronger bound in a sample
with shorter $r_{0}$ and an environment with smaller $\kappa $, wherein the
Keldysh potential is more effective. As Eq. (11) shows, in the same
dielectric screening conditions, heavier excitons with lesser kinetic energy
are more robust with more binding energy and correspondingly smaller radius.
We put in Figure 2 the variation of these exciton features with $\kappa $ in
four ML TMDs, whose inherent chracteristics $\mu _{r}$ and $r_{0}$ are taken
from experiments.$^{32,33}$ Experimentally, the reduction of the exciton
size with environment screening is reported in ref. 57 and the increase of
the exciton binding energy -- ref 58. Moreover, Hsu et al. observe a close
agreement of their findings with the description of the Keldysh potential.$%
^{58}$ As one sees from Fig. 2b, the Keldysh potential also gives for $E_{b}$
in hBN encapsulated ML TMDs the amounts close to experimental ones, though
with slight underestimates.$^{32,33}$

In the first nonlinear approximation, two-pair correlations give rise to the
exciton-exciton interaction. They are mediated by the interaction terms in
Hamiltonian (6) with interacting carriers belonging to two different
electron-hole pairs. We show these correlations in detail in the
Supplementary Note, where one can see the mechanism of various components of
the exciton-exciton interaction. In particular, the interaction between
identical bright excitons and between those from opposite valleys. As
expected, the carrier-carrier interaction within a single valley mediates
the correlations leading to the interaction between excitons in this valley.
Meanwhile, as one can see from Supplementary Fig. 2, the intervalley
exciton-exciton coupling is induced not only by the intervalley
carrier-carrier interactions of all types, but also by the intravalley
electron-hole interactions. As a result, we obtain a Hamiltonian of the
effective exciton-exciton interaction in the form,%
\begin{eqnarray}
\mathcal{H}_{x-x} &=&\frac{1}{2S}\sum\limits_{\mathbf{\mathcal{K}}_{1},%
\mathbf{\mathcal{K}}_{2},\mathbf{Q}}\left\{ \sum\limits_{\Gamma }\left[
U^{d}(\mathbf{Q})+U^{ex}(\mathbf{\mathcal{K}},\mathbf{Q})\right] A_{\Gamma ,%
\mathbf{\mathcal{K}}_{1}+\mathbf{Q}}^{+}A_{\Gamma ,\mathbf{\mathcal{K}}_{2}-%
\mathbf{Q}}^{+}A_{\Gamma ,\mathbf{\mathcal{K}}_{2}}A_{\Gamma ,\mathbf{%
\mathcal{K}}_{1}}\right.  \notag \\
&&+2\sum\limits_{\substack{ \Gamma =\Gamma _{2},\Gamma _{3}  \\ \Gamma
^{\prime }=\Gamma _{5},\Gamma _{6}}}\left[ U^{d}(\mathbf{Q})+U^{ex}(\mathbf{%
\mathcal{K}},\mathbf{Q})\right] A_{\Gamma ,\mathbf{\mathcal{K}}_{1}+\mathbf{Q%
}}^{+}A_{\Gamma ^{\prime },\mathbf{\mathcal{K}}_{2}-\mathbf{Q}}^{+}A_{\Gamma
^{\prime },\mathbf{\mathcal{K}}_{2}}A_{\Gamma ,\mathbf{\mathcal{K}}_{1}} 
\notag \\
&&+2U^{d}(\mathbf{Q})\left[ A_{\Gamma _{2},\mathbf{\mathcal{K}}_{1}+\mathbf{Q%
}}^{+}A_{\Gamma _{3},\mathbf{\mathcal{K}}_{2}-\mathbf{Q}}^{+}A_{\Gamma _{3},%
\mathbf{\mathcal{K}}_{2}}A_{\Gamma _{2},\mathbf{\mathcal{K}}_{1}}+A_{\Gamma
_{5},\mathbf{\mathcal{K}}_{1}+\mathbf{Q}}^{+}A_{\Gamma _{6},\mathbf{\mathcal{%
K}}_{2}-\mathbf{Q}}^{+}A_{\Gamma _{6},\mathbf{\mathcal{K}}_{2}}A_{\Gamma
_{5},\mathbf{\mathcal{K}}_{1}}\right]  \notag \\
&&-\left. 2U^{ex}(\mathbf{\mathcal{K}},\mathbf{Q})\left[ A_{\Gamma _{2},%
\mathbf{\mathcal{K}}_{1}+\mathbf{Q}}^{+}A_{\Gamma _{3},\mathbf{\mathcal{K}}%
_{2}-\mathbf{Q}}^{+}A_{\Gamma _{6},\mathbf{\mathcal{K}}_{2}}A_{\Gamma _{5},%
\mathbf{\mathcal{K}}_{1}}+A_{\Gamma _{5},\mathbf{\mathcal{K}}_{1}+\mathbf{Q}%
}^{+}A_{\Gamma _{6},\mathbf{\mathcal{K}}_{2}-\mathbf{Q}}^{+}A_{\Gamma _{3},%
\mathbf{\mathcal{K}}_{2}}A_{\Gamma _{2},\mathbf{\mathcal{K}}_{1}}\right]
\right\} ,
\end{eqnarray}%
where $U^{d}(\mathbf{Q})$ and $U^{ex}(\mathbf{\mathcal{K}},\mathbf{Q})$ ($%
\mathbf{\mathcal{K}}=\mathbf{\mathcal{K}}_{1}-\mathbf{\mathcal{K}}_{2}$)
denote the direct and exchange interaction energy densities. They are
functions of the Keldysh potential and four envelope functions of two
interacting excitons before and after the interaction, 
\begin{eqnarray}
U^{d}\left( \mathbf{Q}\right) &=&\frac{V_{\mathbf{Q}}}{S^{2}}\left\{ \sum_{%
\mathbf{p}_{1}}\digamma \left( \mathbf{p}_{1}+\alpha \mathbf{Q}\right)
\digamma (\mathbf{p}_{1})^{\ast }\sum_{\mathbf{p}_{2}}\digamma \left( 
\mathbf{p}_{2}-\alpha \mathbf{Q}\right) \digamma \left( \mathbf{p}%
_{2}\right) ^{\ast }\right.  \notag \\
&&+\sum_{\mathbf{p}_{1}}\digamma \left( \mathbf{p}_{1}-\beta \mathbf{Q}%
\right) \digamma (\mathbf{p}_{1})^{\ast }\sum_{\mathbf{p}_{2}}\digamma
\left( \mathbf{p}_{2}+\beta \mathbf{Q}\right) \digamma \left( \mathbf{p}%
_{2}\right) ^{\ast }  \notag \\
&&\left. -2\sum_{\mathbf{p}_{1}}\digamma \left( \mathbf{p}_{1}+\alpha 
\mathbf{Q}\right) \digamma (\mathbf{p}_{1})^{\ast }\sum_{\mathbf{p}%
_{2}}\digamma \left( \mathbf{p}_{2}+\beta \mathbf{Q}\right) \digamma \left( 
\mathbf{p}_{2}\right) ^{\ast }\right\} ,
\end{eqnarray}%
\begin{eqnarray}
U^{ex}\left( \mathbf{\mathcal{K}},\mathbf{Q}\right) &=&-\frac{1}{S^{2}}%
\sum\limits_{\mathbf{p}_{1},\,\mathbf{p}_{2}}V_{p_{1}}\left\{ \digamma
\left( \mathbf{p}_{2}+\alpha \mathbf{Q}\right) \digamma \left( \mathbf{p}%
_{2}\right) ^{\ast }\right.  \notag  \label{e2.36} \\
&&\times \digamma \left( \mathbf{p}_{2}-\mathbf{p}_{1}+\beta \mathbf{%
\mathcal{K+}}\beta \mathbf{Q}\right) \,\digamma \left( \mathbf{p}_{2}-%
\mathbf{p}_{1}+\beta \mathbf{\mathcal{K+}Q}\right) ^{\ast }  \notag \\
&&+\digamma \left( \mathbf{p}_{2}-\beta \mathbf{Q}\right) \digamma \left( 
\mathbf{p}_{2}\right) ^{\ast }\digamma \left( \mathbf{p}_{2}-\mathbf{p}%
_{1}-\alpha \mathbf{\mathcal{K}}-\alpha \mathbf{Q}\right) \digamma \left( 
\mathbf{p}_{2}-\mathbf{p}_{1}-\alpha \mathbf{\mathcal{K-}Q}\right) ^{\ast } 
\notag \\
&&-2\digamma \left( \mathbf{p}_{2}-\mathbf{p}_{1}+\alpha \mathbf{Q}\right)
\digamma \left( \mathbf{p}_{2}\right) ^{\ast }  \notag \\
&&\left. \times \digamma \left( \mathbf{p}_{2}-\mathbf{p}_{1}+\beta \mathbf{%
\mathcal{K+}}\beta \mathbf{Q}\right) \,\digamma \left( \mathbf{p}_{2}-%
\mathbf{p}_{1}+\beta \mathbf{\mathcal{K+}Q}\right) ^{\ast }\right\} .
\end{eqnarray}%
The first, second, and last terms in braces on the right hand side (rhs) of
these equations stand for the energy density of the direct [Eq. (13)] and
exchange [Eq. (14)] exciton-exciton interaction induced by the
electron-electron, hole-hole, and electron-hole interaction, respectively.
The opposite sign of $U^{ex}$ is caused by the exchange of two
carriers-fermions belonging to two interacting excitons. In the case of the
intervalley interaction between the bright excitons, the respective terms
are visualized in Supplementary Fig. 2. Equation (12) formulates the whole
picture of the pairwise interaction between diverse excitons in the model
under consideration, including three types. That is the interaction between
identical excitons, between bright and dark excitons, and between bright
excitons from opposite valleys, presented respectively by the first sum,
second sum, and the last two terms in the braces in rhs of the equation. It
is worth noting that only in terms of excitons valley momenta $\mathcal{H}%
_{x-x}$ has such a relatively compact form as Eqs. (12) -- (14). In terms of
their crystal momenta, each term of $\mathcal{H}_{x-x}$ has its respective
form with the corresponding $U^{d}(\mathbf{Q})$ and $U^{ex}(\mathbf{\mathcal{%
K}},\mathbf{Q})$ [see Supplementary Eqs. (6) -- (10)]. Equation (13) shows
that the direct interaction disappears in the limit of small momentum
transfer and the case of equal electron and hole masses. Therefore this part
matters only at relatively far distances and when one of the exciton
constituents is much heavier than the other. In ML TMDs, the electron and
hole masses are comparable,$^{19,30,36,37}$ so we will put $\alpha \simeq
\beta \simeq 1/2$ and consider the exciton-exciton interaction exclusively
of the exchange nature. Description of the exchange exciton-exciton
interaction is a problematic issue because of its nonlocality.$^{52}$ We
will draw the interaction's qualitative features from basic symmetry
principles and perform an approximate quantitative analysis relying on
calculations for limiting cases. As a result of the Pauli exclusion
principle, the exchange exciton-exciton interaction is short-range,
repulsive between identical excitons and between bright and dark ones. We
see from Supplementary Fig. 1 that a correlated structure of two identical
(or one bright and one dark) excitons incorporate two (or one) couples of
indistinguishable carriers-fermions. The last repel each other at distances,
where their wave functions overlap,$^{59}$ resulting in a repulsive
interaction between excitons. Meantime, Supplementary Fig. 2 shows that all
carriers are distinguishable in a two-exciton structure incorporating
different bright excitons, which can be on equal terms presented as a pair
of dark ones. As $\Gamma _{2}\otimes \Gamma _{3}=\Gamma _{5}\otimes \Gamma
_{6}=\Gamma _{1}$, such a structure is symmetric corresponding to the zero
spin. Symmetric configurations are known to produce attractive forces.$^{59}$
Identical carriers in these structures have opposite spins compensating each
other to the total spin 0, so the attraction is analogous to a chemical
valence bond.$^{34}$

Thus, the exciton-exciton interaction in ML TMDs shares the same dependence
on the interacting excitons' symmetry, or spin, as in 2D or 3D conventional
semiconductors with two simple bands and dipole allowed interband transition.%
$^{49-51}$ It is repulsive in all symmetry combinations of interacting
excitons except for the case they together form a fully symmetric
two-exciton structure (with total spin 0) when the interaction is
attractive. As to the bright excitons, their interaction is repulsive for
parallel spins and attractive for opposite ones.$^{49}$

\subsection*{Intravalley and intervalley exciton interaction potentials}

Let us take a closer look at the intravalley and intervalley interaction
energy. Consider first energy $E_{xx}\left( \Gamma _{2},\mathbf{\mathcal{K}}%
_{1},\mathbf{\mathcal{K}}_{2}\right) $ of two correlated bright $\Gamma _{2}$
excitons with momenta $\mathbf{\mathcal{K}}_{1}$ and $\mathbf{\mathcal{K}}%
_{2}$ in a structure of the type depicted in Supplementary Fig. 1a, d.
Presenting the structure in the zeroth order approximation simply as $%
A_{\Gamma _{2},\mathbf{\mathcal{K}}_{1}}^{+}A_{\Gamma _{2},\mathbf{\mathcal{K%
}}_{2}}^{+}|\,\mathbf{0\,)}\diagup \sqrt{2}$, we get the average of the
exciton Hamiltonian $\mathcal{H}_{x}=\sum\limits_{\Gamma ,\,\mathbf{\mathcal{%
K}}}E_{x}(\Gamma ,\mathbf{\mathcal{K}})A_{\Gamma ,\,\mathbf{\mathcal{K}}%
}^{+}\,A_{\Gamma ,\,\mathbf{\mathcal{K}}}+\mathcal{H}_{x-x}$ over it in the
form of the sum of energies of two excitons and of their interaction, $%
E_{x}\left( \Gamma _{2},\mathbf{\mathcal{K}}_{1}\right) +E_{x}\left( \Gamma
_{2},\mathbf{\mathcal{K}}_{2}\right) +\left[ U_{\Gamma _{2}-\Gamma
_{2}}^{ex}(\mathbf{\mathcal{K}},0)+U_{\Gamma _{2}-\Gamma _{2}}^{ex}(\mathbf{%
\mathcal{K}},-\mathbf{\mathcal{K}})\right] \diagup 2S$. In the expression
for $U_{\Gamma _{2}-\Gamma _{2}}^{ex}(\mathbf{\mathcal{K}},0)$ and $%
U_{\Gamma _{2}-\Gamma _{2}}^{ex}(\mathbf{\mathcal{K}},-\mathbf{\mathcal{K}})$
in terms of excitons' crystal momenta [see Supplementary Eq. (7)], in the
place of Fourier images $V_{k}$ of the Keldysh potential and of exciton wave
functions we insert their Fourier transformations by definition. After some
elementary transfigurations, we get the integral representation of the
intravalley interaction energy of two excitons with crystal momenta $\mathbf{%
k}_{1}$ and $\mathbf{k}_{2}$,%
\begin{eqnarray}
&&\frac{2}{S}\int \!d^{2}r\,\exp [i\alpha \mathbf{kr}]\int
\!d^{2}r_{1}\,f(r_{1})\,f\left( \left\vert \mathbf{r}_{1}+\mathbf{r}%
\right\vert \right)  \notag \\
&&\times \int \!d^{2}r_{2}\,\,f(r_{2})\,f\left( \left\vert \mathbf{r}_{2}-%
\mathbf{r}\right\vert \right) \left\{ V_{K}\left( \left\vert \mathbf{r}_{2}-%
\mathbf{r}\right\vert \right) -V_{K}\left( r\right) \right\}  \notag \\
&\equiv &\frac{1}{S}\int \!d^{2}r\,\exp [i\mathbf{kr}]\,\mathcal{U}_{\Gamma
-\Gamma }(r)
\end{eqnarray}%
where $\mathbf{k\equiv k}_{1}-\mathbf{k}_{2}$. Function of distance $%
\mathcal{U}_{\Gamma -\Gamma }(r)$ ($\Gamma =\Gamma _{2},\Gamma _{3}$)
defines the intravalley interaction potential between identical bright
excitons in the limit of vanishing momentum transfer. For $\mathbf{k}=0$,
the integral of $\mathcal{U}_{\Gamma -\Gamma }(r)$ over 2D space gives us
the intravalley interaction energy in the limit of equal momenta of
interacting excitons, $U_{_{\Gamma -\Gamma }}^{ex}(0,0)\equiv U_{_{\Gamma
-\Gamma }}^{ex}$ ($\Gamma =\Gamma _{2},\Gamma _{3}$). Independent of the
valley position, $\mathcal{U}_{\Gamma -\Gamma }(r)$ is closely similar to
its counterpart in conventional semiconductors: each of its terms is a
product of two two-center integrals met in theory of diatomic molecules.$%
^{52,60}$ The difference is, the carrier-carrier interaction has the form of
the Keldysh potential and the consequent exciton wave function is a
variational one instead of the hydrogenic function.

As to the intervalley interaction energy of different bright excitons, let
us present a symmetric two-exciton structure of the type in Supplementary
Fig. 2 in the form $\left[ A_{\Gamma _{2},\mathbf{\mathcal{K}}%
_{1}}^{+}A_{\Gamma _{3},\mathbf{\mathcal{K}}_{2}}^{+}+A_{\Gamma _{5},\mathbf{%
\mathcal{K}}_{1}}^{+}A_{\Gamma _{6},\mathbf{\mathcal{K}}_{2}}^{+}\right] |\,%
\mathbf{0\,)}\diagup \sqrt{2}$, which diagonalizes the exciton Hamiltonian
average. In terms of excitons' crystal momenta [see Supplementary Eqs. (9),
(10)], the structure energy has the form 
\begin{eqnarray}
&&E_{x}\left( k_{1}\right) +E_{x}\left( k_{2}\right) -\frac{2}{S}U_{\Gamma
_{2}-\Gamma _{3}}^{ex}(\mathbf{k},0)  \notag \\
&&+E_{x}\left( \mathbf{k}_{1}+\mathbf{K-K}^{\prime }\right) +E_{x}\left( 
\mathbf{k}_{2}+\mathbf{K}^{\prime }\mathbf{-K}\right) -\frac{2}{S}U_{\Gamma
_{5}-\Gamma _{6}}^{ex}(\mathbf{k}+2\mathbf{K}-2\mathbf{K}^{\prime },0)
\end{eqnarray}%
where the excitons intervalley interaction energy reads%
\begin{eqnarray}
\frac{1}{S}U_{\Gamma _{2}-\Gamma _{3}}^{ex}(\mathbf{k},0) &=&\frac{1}{S}%
U_{\Gamma _{5}-\Gamma _{6}}^{ex}(\mathbf{k},0)^{\ast }  \notag \\
&=&\frac{2}{S}\int \!d^{2}r\,\exp [i\alpha \mathbf{kr}]\int \int
\!d^{2}r_{1}d^{2}r_{2}\exp \left[ i\alpha \left( \mathbf{K-K}^{\prime
}\right) \left( \mathbf{r}_{1}-\mathbf{r}_{2}\right) \right]  \notag \\
&&\times f(r_{1})\,f(|\mathbf{r}_{1}+\mathbf{r}|)\,f(r_{2})\,f(|\mathbf{r}%
_{2}-\mathbf{r}|)\left[ V_{K}\left( |\mathbf{r}_{2}-\mathbf{r}|\right)
-V_{K}(r)\right]  \notag \\
&\equiv &-\frac{1}{S}\int \!d^{2}r\,\exp [i\mathbf{kr}]\,\mathcal{U}_{\Gamma
_{2}-\Gamma _{3}}(r)
\end{eqnarray}%
Here $\mathcal{U}_{\Gamma _{2}-\Gamma _{3}}(r)$ is an attractive interaction
potential that outlines the interaction potential between $\Gamma _{2}$ and $%
\Gamma _{3}$ excitons in the limit of vanishing momentum transfer. Similarly
to $U_{_{\Gamma -\Gamma }}^{ex}$ ($\Gamma =\Gamma _{2},\Gamma _{3}$), we
quantify the intervalley interaction potential by the value of the integral
of $\left\vert \mathcal{U}_{\Gamma _{2}-\Gamma _{3}}(r)\right\vert $ over 2D
space that we denote by $U_{\Gamma _{2}-\Gamma _{3}}^{ex}$. We will refer to
the quantity as the intervalley interaction energy (in the limit of equal
momenta). Comparing Eqs. (17) and (15) we see that the interaction between
distant $\Gamma _{2}$ and $\Gamma _{3}$ excitons in the momentum space is
described by an oscillating exponent in the inner two-center integrals (over 
$\mathbf{r}_{1}$ and $\mathbf{r}_{2}$). From Eq. (5) and the form of the
exciton wave function, we have the oscillation frequency $2\pi /(3\sqrt{3}%
)a_{x}/a$, which roughly varies between 3.5 and 6.5 for $a\sim 3$ \AA\ and $%
a_{x}$ in the range $9-17$ \AA\ (see Fig. 2a). Consequently, the intervalley
interaction potential is much weaker compared to its intravalley
counterpart. Essentially, the potential depends substantially on the exciton
extent defining the overlap degree of the wave functions of interacting
excitons from $K$ and $K^{\prime }$ valleys in the momentum space. As
expected, the smaller the exciton radius (the more extended the exciton wave
function $\digamma (\mathbf{k})$) is, the stronger the intervalley
interaction potential.

\begin{figure}[tbp]
\centering
\includegraphics[width=4in,height=3.09in,keepaspectratio]{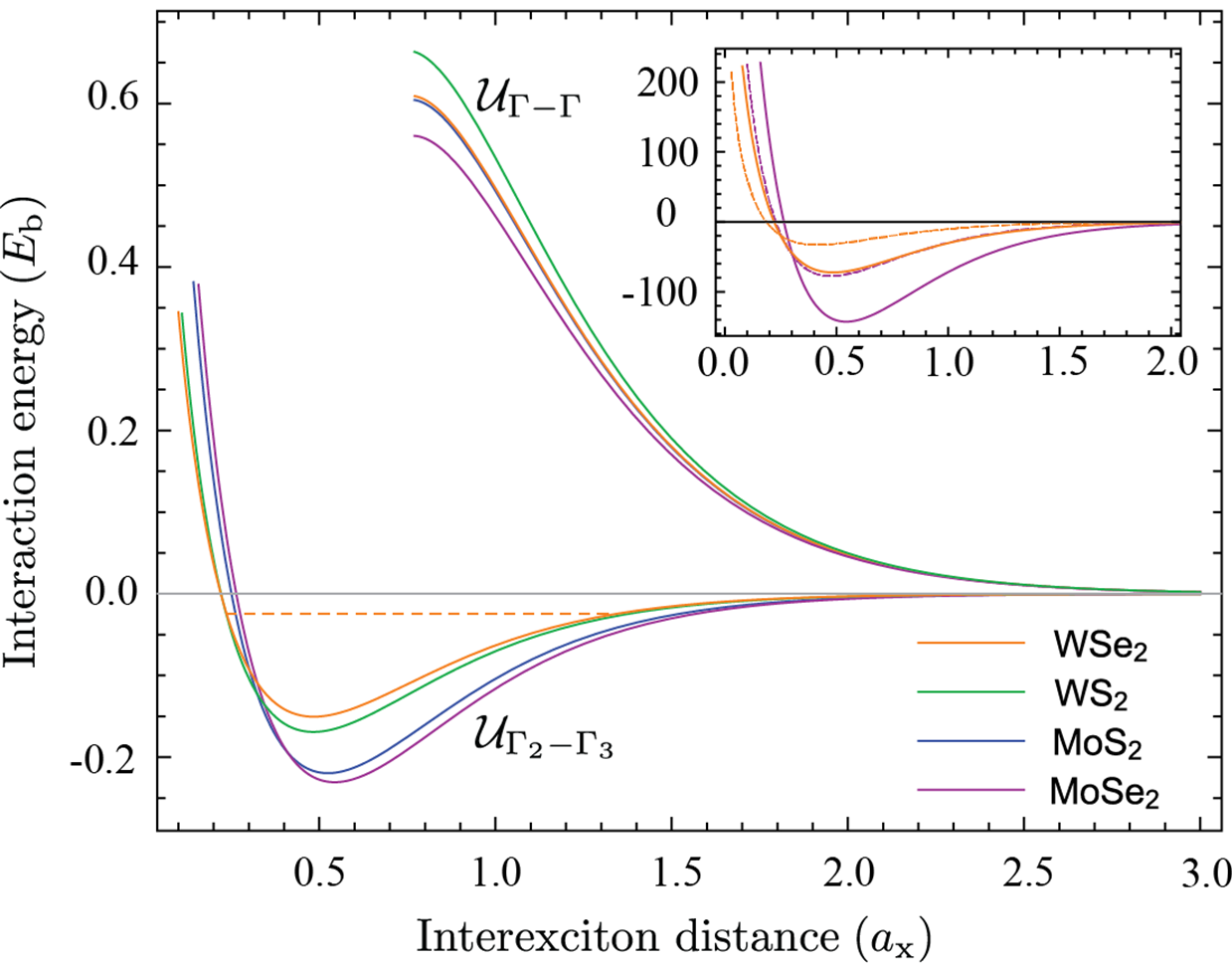}
\caption{Intravalley $\mathcal{U}_{\Gamma -\Gamma }$ ($\Gamma =\Gamma
_{2},\Gamma _{3}$) and intervalley exciton interaction potential $\mathcal{U}%
_{\Gamma _{2}-\Gamma _{3}}$ in the limit of vanishing momentum transfer in
different freestanding ML TMDs. The dashed line inside the intervalley
potential in ML WSe$_{2}$ depicts the approximate position of the
intervalley biexciton energy. We put lattice constant $a=3.3$ \AA\ for ML
MoSe$_{2}$ and WSe$_{2}$, and $a=3.2$ \AA\ for ML MoS$_{2}$ and WS$_{2}$.$%
^{37}$ Inset: The intervalley interaction potential in ML MoSe$_{2}$ and WSe$%
_{2}$ in vacuum (solid lines) and hBN encapsulation (dashed lines) in the
unit of millielectronvolt.}
\label{fig:F3}
\end{figure}

With the particular experimental value of $\mu _{r}$ and $r_{0}$\ for
different ML TMDs,$^{33}$ the exciton radius, binding energy, and
interaction energies $U_{\Gamma -\Gamma }^{ex}$ and $U_{\Gamma _{2}-\Gamma
_{3}}^{ex}$\ obtained for freestanding ML TMDs are shown in the upper part
of Table 1, and potentials $\mathcal{U}_{\Gamma -\Gamma }(r)$ and $\mathcal{U%
}_{\Gamma _{2}-\Gamma _{3}}(r)$ as functions of relative distance $r/a_{x}$
-- in Fig. 3. The interaction potentials (energies) computed for a ML TMD
are expressed in the unit of $E_{b}$ ($E_{b}a_{x}^{2}$) found for the
material. As the electron-hole interaction energy [see Eq. (11)], the
intravalley interaction potential in its absolute quantity weakens with
increasing ratio $r_{0}/\kappa a_{x}$. For $\kappa =1$ this trend is seen
from amounts of $r_{0},a_{x}$ and $U_{_{\Gamma -\Gamma }}^{ex}$ in the upper
part of Table 1 and Fig. 3. Meanwhile, the potential shape remains the same
for all ML TMDs with a repulsive wall at small distances and an exponential
fall at larger ones. These features are characteristic of the interaction
between excitons,$^{52,60}$ resembling that between atoms in diatomic
molecules.$^{34}$ Further, for freestanding ML TMDs, the intravalley
interaction strength is somewhat above $2E_{b}\,a_{x}^{2}$. On the scale $%
\sim 2E_{b}\,a_{x}^{2}$ it has been guesstimated in ref. 9 for ML WS$_{2}$.
If in Eq. (15) replace $V_{K}(r)$ by the Coulomb potential, the integral
gives the known result $U^{ex}|_{C}=8\pi E_{b}\,a_{x}^{2}\left( 1-315\pi
^{2}/2^{12}\right) \approx 6E_{b}\,a_{x}^{2}$.$^{48,51}$ Thus, reduced
screening makes the relative exciton-exciton interaction energy nearly
thrice less compared to its amount in the limit of Coulomb potential $%
V_{C}(r)$. The exciton-exciton interaction terms mediated respectively by
the electron-hole, electron-electron, and hole-hole interactions become more
with the reduced screening. However, their magnitudes are closer to each
other, so their difference decreases, involving lesser exciton-exciton
interaction relative to the exciton binding energy. The result means high
stability of valley excitons relative to their pairwise interaction and a
wide valid range of the low-density limit. We should note that as $E_{b}$ in
ML TMDs is by orders more than in conventional semiconductors, in its
absolute quantity, the exciton-exciton interaction in the former is
enhanced, compared to that in the latter.$^{9,24}$

\begin{table}[t]
\caption{Experimental values of input variables in different ML TMDs and
resultant calculated exciton binding energy and radius, intravalley and
intervalley exciton interaction energies, and biexciton binding energy.
Available appropriate experimental measurements of the exciton and biexciton
binding energies are shown in parentheses for comparison. }
\label{Table1}\centering
\centering
\begin{tabular}{p{1.29in}p{1.29in}p{1.29in}p{1.29in}p{1.29in}}
\hline\hline
&  &  &  &  \\ 
& MoSe$_{2}$ & MoS$_{2}$ & WS$_{2}$ & WSe$_{2}$ \\ \hline
&  &  &  &  \\ 
$\mu _{r}$ ($m_{0}$)$^{33}$ & 0.350 & 0.275 & 0.175 & 0.200 \\[5pt] 
$r_{0}$ (\AA )$^{33}$ & 39 & 34 & 34 & 45 \\[5pt] \hline
&  &  &  &  \\ 
&  & Freestanding &  &  \\ \hline
&  &  &  &  \\ 
$E_{b}$ (meV) & 620 & 644 & 566 & 481 \\[5pt] 
$a_{x}$ (\AA ) & 8.38 & 8.96 & 11.48 & 12.11 \\[5pt] 
$U_{\Gamma -\Gamma }^{ex}$ ($E_{b}a_{x}^{2}$) & 2.043 & 2.173 & 2.341 & 2.187
\\[5pt] 
$U_{\Gamma _{2}-\Gamma _{3}}^{ex}$ ($E_{b}a_{x}^{2}$) & 0.952 & 0.863 & 0.602
& 0.539 \\[5pt] 
$\mathcal{E}_{0}$ (meV) & 64.9 & 53.4 (60$^{12}$) & 18.5 & 12.9 \\%
[7pt] \hline
&  &  &  &  \\ 
&  & hBN encapsulated &  &  \\ \hline
&  &  &  &  \\ 
$\kappa ^{33}$ & 4.4 & 4.45 & 4.35 & 4.5 \\[5pt] 
$E_{b}$ (meV) & 226 (231$^{33}$) & 215 (221$^{33}$) & 174 (180$^{33}$) & 157
(167$^{32}$) \\[5pt] 
$a_{x}$ (\AA ) & 10.38 & 11.53 & 15.39 & 15.69 \\[5pt] 
$U_{\Gamma -\Gamma }^{ex}$ ($E_{b}a_{x}^{2}$) & 3.355 & 3.584 & 3.826 & 3.620
\\[5pt] 
$U_{\Gamma _{2}-\Gamma _{3}}^{ex}$ ($E_{b}a_{x}^{2}$) & 1.177 & 0.996 & 0.642
& 0.615 \\[5pt] 
$\mathcal{E}_{0}$ (meV) & 24.0 (21$^{27}$) & 13.2 & 1.5 & 1.4 (16 -- 17$%
^{18} $) \\[5pt] \hline\hline
\end{tabular}%
\end{table}

As to the intervalley interaction, the weakening with increasing exciton
size can be seen from Table 1 and Fig. 3. The intervalley interaction energy 
$U_{\Gamma _{2}-\Gamma _{3}}^{ex}$ is one half of the intravalley
counterpart $U_{\Gamma -\Gamma }^{ex}$ in freestanding ML MoSe$_{2}$ with
the smallest exciton, but just a fourth of $U_{\Gamma -\Gamma }^{ex}$ in ML
WSe$_{2}$ with the largest one. Figure 3 shows that $\mathcal{U}_{\Gamma
_{2}-\Gamma _{3}}(r)$ is very short-range with its depth increasing with the
decrease of $a_{x}$. Meantime, the potential width $a_{w}$ (width at half
minimum) decreases in absolute quantity remaining in a narrow interval $%
0.65a_{x}-0.67a_{x}$ in freestanding ML TMDs. In the presence of environment
dielectric screening, the potential width increases with $a_{x}$ in absolute
quantity though it slightly decreases in the unit of $a_{x}$. Examinations
show that even for the most strong potential $\mathcal{U}_{\Gamma
_{2}-\Gamma _{3}}^{m}(r)$ in freestanding ML MoSe$_{2}$, its average value
(the value at half minimum) meets the inequality $\left\vert \overline{%
\mathcal{U}_{\Gamma _{2}-\Gamma _{3}}^{m}(r)}\right\vert \ll 2\diagup \mu
_{x}a_{w}^{2}$. Thus for any possible value of the input variables, the
intervalley interaction potential can be considered as a perturbation$^{34}$
to the free relative motion of $\Gamma _{2}$ and $\Gamma _{3}$ excitons in
their correlated symmetric structures. Clearly, the shallower the potential
is, the better the perturbation criterion is fulfilled. To the detail, the
criterion inequality corresponds to the ratio of 1 to 5, 1 to 6, and 1 to 9
for freestanding ML MoSe$_{2}$, MoS$_{2}$, and both members of MoX$_{2}$
subgroup, respectively, while in an hBN encapsulation it is correspondingly
1 to 7,\ 1 to 8, and 1 to 14.

\subsection*{Intervalley biexciton}

Let us consider a superposition of correlated symmetric structures of two
bright excitons from opposite valleys having a definite valley momentum $%
\mathbf{\mathcal{P}}$ 
\begin{eqnarray}
|\Gamma _{1},\mathbf{\mathcal{P}}\,) &=&\frac{1}{\sqrt{2S}}\sum\limits_{%
\mathbf{\mathcal{K}}}\Psi (\mathbf{\mathcal{K}})\left[ A_{\Gamma _{2},\,%
\mathbf{\mathcal{K}+\mathcal{P}}/2\,}^{+}A_{\Gamma _{3},\mathbf{-\mathcal{K}+%
\mathcal{P}}/2}^{+}+A_{\Gamma _{5},\mathbf{\mathcal{K}+\mathcal{P}}%
/2\,}^{+}A_{\Gamma _{6},\mathbf{-\mathcal{K}+\mathcal{P}}/2}^{+}\right]
\,|\,0\,)  \notag \\
&\equiv &B_{\mathbf{\mathcal{P}}}^{+}\,|\,0\,)
\end{eqnarray}%
It is straightforward to examine, that this two-exciton entity, whose valley
momentum equals its crystal momentum, is an eigenstate of Hamiltonian $%
\mathcal{H}_{x}$ of the exciton system with effective exciton-exciton
interaction,%
\begin{equation}
\mathcal{H}_{x}|\Gamma _{1},\mathbf{\mathcal{P}}\,)=E_{xx}(\mathcal{P}%
)|\Gamma _{1},\mathbf{\mathcal{P}}\,)
\end{equation}%
with energy $E_{xx}(\mathbf{\mathcal{P}})=2E_{x}+\mathbf{\mathcal{P}}%
^{2}/4\mu _{x}+\mathcal{E}_{xx}$ and the envelope function $\Psi (\mathbf{%
\mathcal{K}})$ obeying the equation 
\begin{equation}
\frac{\mathcal{K}^{2}}{\mu _{x}}\Psi (\mathbf{\mathcal{K}})-\frac{2}{S}%
\sum\limits_{\mathbf{\mathcal{Q}}}U_{\Gamma _{2}-\Gamma _{3}}^{ex}(2\mathbf{%
\mathcal{K}},\mathbf{\mathcal{Q}})\Psi (\mathbf{\mathcal{K}}+\mathbf{%
\mathcal{Q}})=\mathcal{E}_{xx}\Psi (\mathbf{\mathcal{K}})
\end{equation}%
We see, that $|\Gamma _{1},\mathbf{\mathcal{P}}\,)$ is a correlated
two-exciton entity with total mass $2\mu _{x}$ and reduced mass $\mu _{x}/2$%
, whose energy includes the kinetic energy of its free motion as a whole and
internal energy $\mathcal{E}_{xx}$ of the excitons relative motion in the
field of their mutual attractive interaction. The internal energy is
positive for the scattering states and negative for bound states with
binding energy $\mathcal{E}_{b}=-\mathcal{E}_{b,xx}>0$. We call $|\Gamma
_{1},\mathbf{\mathcal{P}}\,)$ the intervalley biexciton in the broad sense
of the word though conventionally it is used to refer to the bound state
only. As a nonlocal function of two vectors-variables, $U_{\Gamma
_{2}-\Gamma _{3}}^{ex}(2\mathbf{\mathcal{K}},\mathbf{\mathcal{Q}})$ is
presented in the real space by a nonlocal potential $\mathcal{U}_{\Gamma
_{2}-\Gamma _{3}}(r,r^{\prime })^{60}$ and Eq. (20) is therefore an
integrodifferential equation. It cannot be analyzed with usual methods of
nonlinear dynamics and bound to be reduced by approximations. The approach
proposed in ref. 52, which consists in expanding the exchange interaction
energy density into a series of powers of $\alpha $, can be applied.
However, with $\alpha \approx \beta \approx 1/2$, one has to retain a large
number of the series terms yielding a high order differential equation.
Dealing with such an approximate solution of Eq. (20) is itself a demanding
issue that is beyond the scope of this paper. We note only that i) $\mathcal{%
U}_{\Gamma _{2}-\Gamma _{3}}(r)$ comes from the main zero-order term of the
mentioned series, ii) the intervalley interaction energy is that of the real
nonlocal potential, $U_{\Gamma _{2}-\Gamma _{3}}^{ex}=\int d^{2}r\left\vert 
\mathcal{U}_{\Gamma _{2}-\Gamma _{3}}(r)\right\vert =\int
d^{2}r\,d^{2}r^{\prime }\left\vert \mathcal{U}_{\Gamma _{2}-\Gamma
_{3}}(r,r^{\prime })\right\vert $, and iii) $\mathcal{U}_{\Gamma _{2}-\Gamma
_{3}}(r)$ in combination with the first and second-order terms of the series
corresponds to an equivalent energy-dependent local potential of the same
range ($a_{w}$), which is slightly deeper with minimum shifted towards a
larger distance. Therefore binding energy $\mathcal{E}_{0}$ of the bound
state supported by $\mathcal{U}_{\Gamma _{2}-\Gamma _{3}}(r)$ can be
considered a lower bound for the biexciton binding energy, i.e. the minimal
amount $\mathcal{E}_{b}$ can have for a set of input variables values, $%
\mathcal{E}_{0}\lesssim \mathcal{E}_{b}$. Because we can acquire $\mathcal{E}%
_{0}$ just approximately as one can see later, it indicates a narrow range,
wherein the biexciton binding energy might be. Thus we will refer to this
quantity approximately as biexciton binding energy. Presenting the\ envelope
function of the bound state as $\psi (\mathbf{r})=R(r)\exp [im\varphi ]/2\pi 
$ with $m$ an integer, we have the equation for its radial part,

\begin{equation}
-\frac{1}{\mu _{x}}\frac{1}{r}\frac{d}{dr}\left[ r\frac{d}{dr}R(r)\right] +%
\left[ \mathcal{U}_{\Gamma _{2}-\Gamma _{3}}(r)+\frac{m^{2}}{\mu _{x}r^{2}}%
\right] R(r)=-\mathcal{E}_{0}\,R(r)
\end{equation}%
Detailed checking shows that already for $m=1$ the centrifugal term
predominates\ in strength over $\mathcal{U}_{\Gamma _{2}-\Gamma _{3}}^{m}(r)$%
. Hence for any realistic amount of input parameters, Eq. (21) with a
repulsive effective potential has no negative solution. Thus the intervalley
interaction potential can support only bound states with $m=0$. Moreover,
our calculations using Eq. (30) in ref. 61 for the number of such bound
states in a 2D potential show that for realistic amounts of the input
parameters, $\mathcal{U}_{\Gamma _{2}-\Gamma _{3}}(r)$ can host only one
bound state. Its binding energy can be estimated in a perturbation theory
manner as shown by Landau and Lifshitz,$^{34}$ 
\begin{equation}
\mathcal{E}_{0}\sim \frac{2}{\mu _{x}a_{w}^{2}}\exp \left\{ -\frac{2}{\mu
_{x}}\frac{2\pi }{U_{\Gamma _{2}-\Gamma _{3}}^{ex}}\right\}
\end{equation}

\begin{figure}[t]
\centering
\includegraphics[width=3.6in,height=2.69in,keepaspectratio]{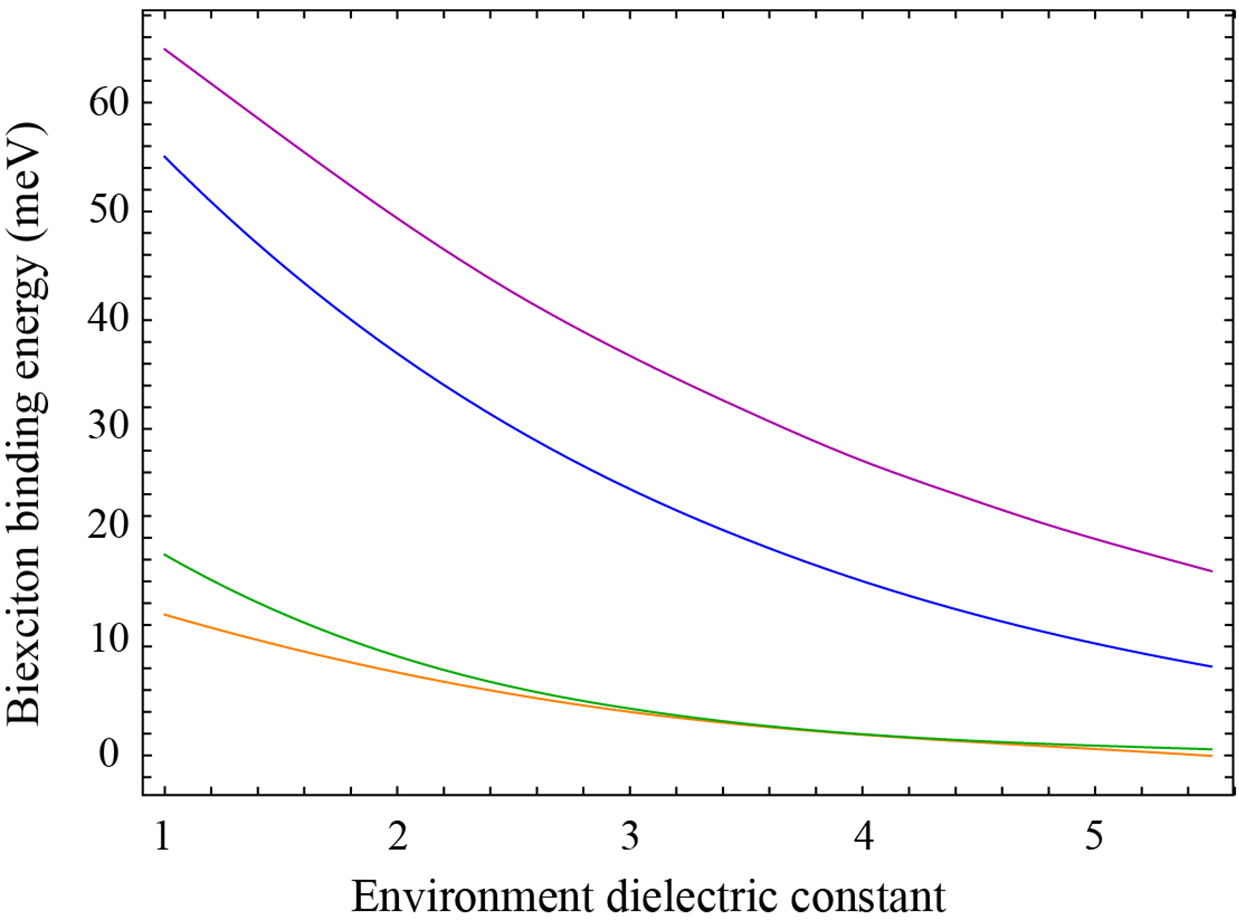}
\caption{Variation of the intervalley biexciton binding energy with
environment screening in four ML TMDs. The lines come from the interpolation
between multiple points calculated with the use of Eqs. (22) and (17). As in
Figs. 2 and 3, the orange line is for ML WSe$_{2}$, green line -- ML WS$_{2}$%
, blue line -- ML MoS$_{2}$, and purple line -- ML MoSe$_{2}$.}
\label{fig:F4}
\end{figure}

We see an exponential increase of the biexciton binding energy with the
exciton mass and intervalley interaction energy $U_{\Gamma _{2}-\Gamma
_{3}}^{ex}$ that itself rises with increasing mass and decreasing screening.
Consequently, $\mathcal{E}_{0}$ is sensitive to any variation of the input
variables, which might help explaining disagreement between the reported
experimental measurements.$^{11-18}$ The sensitivity is most relevant to the
exciton mass, which enters the exponent's degree in Eq. (22), and is one of
the variables determining $U_{\Gamma _{2}-\Gamma _{3}}^{ex}$. Therefore the
biexciton binding energy in ML MoX$_{2}$ with heavier excitons is much more
than in ML WX$_{2}$. One can see from Table 1 that the ratio between the
amounts of $\mathcal{E}_{0}$ in two groups is several times in freestanding
samples. We note that the number we obtain for freestanding ML MoS$_{2}$ is
near the measurement of Sie et al.$^{12}$ To our knowledge, experimental
data of the biexciton binding energy in remaining ML TMDs in a vacuum is not
available. As to the other inherent parameter, the dependence of $\mathcal{E}%
_{0}$ on $r_{0}$\ is through $a_{x}$, whereon $U_{\Gamma _{2}-\Gamma
_{3}}^{ex}$\ strongly depends. We see from Fig. 2a that in any environment,
the exciton is largest in ML WSe$_{2}$ with longest $r_{0}$, though $\mu _{r}
$ in ML WS$_{2}$ is a little lighter. The exciton extent determines the
intervalley interaction strength, so the biexciton binding energy in ML WSe$%
_{2}$ is least, as seen from Fig. 4, where we put the variation of $\mathcal{%
E}_{0}$\ with $\kappa $ in all four ML TMDs. In light of our results for ML
WS$_{2}$ and WSe$_{2}$ shown in the figure, it is unlikely the intervalley
biexciton binding energy in the range of 45--65 meV, which has been deduced
from observed resonances in early experiments on SiO$_{2}$ substrated ML WSe$%
_{2}$ and WS$_{2}$.$^{13-15}$ Doubts have been raised recently about the
nature of those resonances with different mechanisms suggested for their
origin.$^{8.62}$ As one can see from Fig. 1, in ML WX$_{2}$ the energy of
the spin-dark excitons is below that of the bright ones. Therefore the part
of their optical spectra below the exciton resonance is much richer than in
ML MoX$_{2}$. The physical origin of different experimentally observed peaks
in the low-energy part of spectra of ML WX$_{2}$ is still not clear,$^{8}$
so their misinterpretation seems a common practice. Encapsulation by hBN
flakes often used lately$^{7,27,32,33}$ considerably improves the spectra by
reducing excitonic linewidths to $\sim $\ 2--4 meV at 4 K [7]. The marked
increase of $a_{x}$ at $\kappa \sim 4.5$ (see Table 1) leads to a sharp
decline of the intervalley interaction (see Fig. 3, the Inset) involving a
striking decrease of the biexciton binding energy. We find that $\mathcal{E}%
_{0}$ goes down to the range of 24 meV in hBN encapsulated ML MoSe$_{2}$,
which agrees with the recent measurement of Yong et al.$^{27}$ In hBN
encapsulated ML WX$_{2}$, the exciton radius becomes about five times larger
than the lattice constant. The oscillating factor with a frequency of about
6 in rhs of Eq. (17) severely diminishes the intervalley interaction
potential. Quantum mechanically speaking, the exciton wave functions at the $%
K$\ and $K^{\prime }$ valleys become so small and localized in the momentum
space that they can hardly overlap. As a result, $\mathcal{E}_{0}$
dramatically falls off to values less than excitonic linewidths. In this
connection, an amount in the range 16--17 meV reported for the biexciton
binding energy in hBN encapsulated ML WSe$_{2}$ [18] appears to be a
misinterpretation. It is not convincing that the authors claim about the
agreement of their measurement with those of refs. 16 and 17, and also with
theoretical results of refs. 19--22. First, experiments in refs. 16 and 17
are on sapphire substrates, and in the latter, the sample is ML MoSe$_{2}$.
Taking high-frequency dielectric constant of sapphire substrates $%
\varepsilon _{b}=9.3$,$^{63}$ we get numbers about 19 meV and 0.4 meV for $%
\mathcal{E}_{0}$ in ML MoSe$_{2}$ and WSe$_{2}$, respectively. The former
agrees with the experimental measurement of Hao et al.,$^{16}$ giving
another example, that Eq. (22) provides reasonable judgment of $\mathcal{E}%
_{0}$ in ML MoX$_{2}$. Meantime, the obtained number for ML WSe$_{2}$ is
even smaller than the smallest ($3\pm 0.5$ meV), not mentioning the largest (%
$18\pm 0.5$ meV), among three values for $\mathcal{E}_{0}$ the authors of
ref. 17 infer from their time-resolved differential absorption data. For $%
\kappa \approx 5.15$, potential $\mathcal{U}_{\Gamma _{2}-\Gamma _{3}}(r)$
is still shallower than it is in the case of $\kappa =4.5$ shown in the
Inset of Fig. 3, with an average value of about $-15$ meV. The potential
fine meets the perturbation criterion, so estimation by Eq. 22 is credible.

Concerning theoretical works on the biexciton,$^{19-23}$ we should note the
following. The starting intervalley biexciton model in these works is the
same as ours. That is a 2D two-pair structure with indentical carriers
having opposite spins and the Keldysh carrier-carrier interaction. However,
the fact that the two electron-hole pairs forming the intervalley biexciton
come from different edges of BZ has not been taken into consideration.
Besides, the exchange interaction that is the primary part of the
interaction between excitons has been coped with inadequately. Further, with
the Keldysh carrier-carrier interaction, the relationship between the
exciton binding energy and radius is determined by its reduced mass and
dielectric parameters, as described by Eq. (11). By assuming the hydrogenic
model relationship $E_{b}a_{x}=e^{2}/\varepsilon _{0}\varepsilon $, the
authors exclude the exciton mass from their examinations. As a result, their
biexciton binding energy depends only on the screening length and
electron-hole mass ratio. With values of the last differing not much, the
obtained amounts for $\mathcal{E}_{0}$ in different ML TMDs are close to
each other.$^{19-22}$ Overall, that seems the biexciton model considered in
those works has little in common with the intervalley biexciton in a real ML
TMD.

Undoubtedly, our theoretical model involving approximations contains
inaccuracies, and the used experimental measurements of input parameters
entail uncertainties. In connection with the sensitivity of the biexciton
binding energy to the input variables, they might bring about considerable
uncertainty of the result for $\mathcal{E}_{0}$. This concerns first the
perturbation theory estimate in the form of Eq. (22). The better potential $%
\mathcal{U}_{\Gamma _{2}-\Gamma _{3}}(r)$ fulfills the perturbation
criterion, the closer $\mathcal{E}_{0}$ is to the exact value. Details on
the fulfillment in different ML TMDs listed earlier show that the
approximation is good for the ML WX$_{2}$ in any environment and ML MoX$_{2}$
in the presence of environment screening. For freestanding ML MoSe$_{2}$, it
is a rather crude approximation needing further improvement. Secondly, our
model relies on the exciton effective mass description and Keldysh form of
the carrier-carrier interaction leading to Eq. (11). By comparing the amount
of $E_{b}$ following from the equation and its experimental value in hBN
encapsulated ML TMDs (see Fig. 2b and Table 1, the lower part), we see that
Eq. (11) slightly underestimates $E_{b}$, by 5 -- 10 meV. The difference
between the two amounts is most (about six percent) for ML WSe$_{2}$. By
fitting Eq. (11) to 167 meV, we get either $\mu _{r}=0.21m_{0}$, taking into
account the experimental uncertainty pointed out in ref. 32, and $r_{0}=42$ 
\AA , or $\mu _{r}=0.22m_{0}$, as suggested by the authors' group earlier in
ref. 57, and $r_{0}=44$ \AA . The two alternatives yield $\mathcal{E}_{0}$
approximately 19 meV in a vacuum and 2.5 meV in an hBN encapsulation. The
relative change in both cases is sizeable, but in the latter, it does not
change the fact that the biexciton is hard to observe in hBN encapsulated ML
WSe$_{2}$, and in general, in an environment with $\kappa >3$ as one can see
from Fig. 4. Thirdly, from Eq. (15) on, our computations are carried out for 
$\alpha =\beta $ yielding $U^{d}=0$. With the difference between electron
and hole masses taken into consideration, the upper line in (16) gains an
additional term $U^{d}(\mathbf{k}_{1}-\mathbf{k}_{2})$, which corresponds to
a local potential $\mathcal{U}_{\Gamma _{2}-\Gamma _{3}}^{d}(r)$ in the
coordinate space. The direct exciton interaction potential has a
considerable positive value near $r\sim 0$,$^{60}$ then takes minor negative
values adding an insignificant amount to potential $\mathcal{U}_{\Gamma
_{2}-\Gamma _{3}}(r)$. As $\int d^{2}r\mathcal{U}_{\Gamma _{2}-\Gamma
_{3}}^{d}(r)=U_{\Gamma _{2}-\Gamma _{3}}^{d}(0)=0$, it makes no contribution
to intervalley interaction energy $U_{\Gamma _{2}-\Gamma _{3}}^{ex}$. In
this way, with the little difference between $\mu _{e}$ and $\mu _{h}$ in ML
TMDs,$^{19,30,36,37}$ neglecting the direct part of the exciton-exciton
interaction is acceptable.

\section*{DISCUSSION}

We have presented a theoretical model for the system of diverse ground state
excitons in ML TMDs with their effective pairwise interaction in the
low-density limit. We make use of a group theoretical classification scheme
for the band states and related excitons, where each of them is notated by a
one-dimensional irreducible representation of the Abelian group $C_{3h}$,
the wave vector group at the $K $ and $K^{\prime }$ valleys. We limit
ourselves to a simplified band structure of ML TMDs with a direct two-band
scheme at each valley, yielding four exciton symmetry states, two bright and
two dark ones. Analogous states the exciton has in conventional 2D and 3D
semiconductors with twofold spin-degenerate bands and dipole-allowed
interband transition. We find that qualitatively, excitons in ML TMDs
interact with each other in the same way as their conventional counterparts.
That is, the character of the interaction between excitons with $\Gamma $
and $\Gamma ^{\prime }$ notations is defined by the product $\Gamma \otimes
\Gamma ^{\prime }$ representing the two-exciton correlated structure they
together form. It is repulsive for $\Gamma \otimes \Gamma ^{\prime }\neq
\Gamma _{1}$, corresponding to a nonzero excitons total spin, and attractive
in the only symmetric one, $\Gamma \otimes \Gamma ^{\prime }=\Gamma _{1}$,
corresponding to the spin $0$. Concerning the bright excitons, their mutual
interaction is repulsive for parallel spins and attractive for opposite
ones. The distinction of excitons and their pairwise interaction in ML TMDs
is due to the materials' particular band structure and reduced screening in
the form of the Keldysh carrier-carrier interaction. The former drives
excitons with opposite spins residing at inequivalent valleys distant in the
momentum space. In this way, we find in ML TMDs the repulsive intravalley
and attractive intervalley exciton-exciton interaction. The latter,
naturally depending on the overlap degree of the exciton wave functions at
two valleys, supports the intervalley biexciton formation. With the Keldysh
form of the carrier-carrier interaction, the exciton radius determining the
wave function extent is the variational parameter.

Quantitatively, we have established an analytical relationship of
variational parameter $a_{x}$, and the corresponding exciton binding energy,
with the exciton reduced mass and the sample and environment dielectric
characteristics. The latter are thereby the input variables determining the
former as primary features of the exciton, and also the exciton-exciton
interaction and the intervalley biexciton binding energy. We have acquired
the intervalley interaction potential as a function of the interexciton
distance, showing its explicit dependence on the exciton radius. We find
that for realistic values of the input variables, the intervalley
interaction potential turns out to be sufficiently weak, permitting us to
estimate biexciton binding energy $\mathcal{E}_{0}$ in a perturbation theory
manner. In this way, we obtain its semianalytical dependence on the exciton
mass and the sample and environment dielectric parameters. We notice that $%
\mathcal{E}_{0}$ is sensitive to every input variable, especially the
exciton mass. We find that in a vacuum, $\mathcal{E}_{0}$ in
molybdenum-based MLs with heavier excitons is several times larger than in
tungsten-based ones, and the ratio rises to about an order in the presence
of environment screening. The amounts of $\mathcal{E}_{0}$ we estimate for
freestanding ML MoS$_{2}$, and also sapphire substrated and hBN encapsulated
ML MoSe$_{2}$, well agree with available relevant experimental measurements.
Meantime, our estimation for ML WSe$_{2}$ in those conditions gives values
very small compared to two appropriate experimental reports. From our
perspective, this might be connected with misclassifications of the observed
experimental spectra.

The semianalytical relationship established between the exciton and
biexciton binding energy with environment dielectric constant might be used
for adjusting the exciton and biexciton feature of different ML TMDs in
future optoelectronic applications. Further, from our symmetry-dependent
exciton Hamiltonian, a system of Heisenberg equations of motion can be
derived. Such a system would be a baseline for research on valley selective
nonlinear effects in an ML TMD coherently driven near the exciton resonance.
In this connection, it is worthwhile pointing out the applicability of the
presented model. It describes the coherent dynamics dominating the initial
stages of optical experiments$^{64}$ in the first nonlinear regime. Created
exciton polarization has the phase of the exciting field, and its inherent
part has not appeared yet. The interaction with the carriers, phonons, etc.,
available in the sample, whose concentration is assumed small, causes a weak
dephasing of the exciton polarization resulting in a slight reduction of its
coupling with field and an energy shift of the exciton resonance.$^{49}$ The
model is inapplicable to conditions when the created exciton polarization
becomes incoherent, or under an above gap excitation, when the resulting
excited state population is a mixture of bound excitons and electron-hole
plasma. The presence of excess carriers at a moderate density can
considerably affect the exciton, its interaction with the light and each
other.$^{65}$ At an intermediate density, the interaction of excitons with
bound and unbound charged excitons (trions) can dress them into
exciton-polarons.$^{66}$ However, the description of such effects is beyond
the scope of the paper.

\section*{Acknowledgement}

N.T.P. acknowledges funding from JSPS KAKENHI Grant Number 19K14638. H.N.C.
and V.A.O. acknowledge support through Theme 01-3-1137-2019/2023
\textquotedblleft Theory of Complex Systems and Advanced
Materials\textquotedblright\ of Joint Institute for Nuclear Research.

\end{document}


\begin{center}
{\Large \textbf{Supplemetary Note. \\
Derivation of the exciton-exciton interaction\ }}
\end{center}


To consider two-pair correlations, we insert number operator $N$ into the
carrier-carrier interaction terms of Hamiltonian $\mathcal{\mathcal{H}}_{eh}$
(Eq. (6) in the main text) two times, once to the creation operators and
once to the annihilation ones. Let us take for example the first of these
terms describing the interaction between electrons at the $K$ valley. From
the structure of Eq. (9) in the main text we see, that the interacting
electrons can come into correlations with either two holes from the same $K$
valley, two holes from the $K^{\prime }$ valley, or one hole from the $K$
valley and the other from the $K^{\prime }$ valley. These processes are
described respectively by the following operator products,%
\begin{eqnarray}
&&\sum\limits_{\mathbf{q},\,\mathbf{p}_{1},\,\mathbf{p}_{2}}V_{q}\,{\large e}%
_{\mathbf{\Gamma }_{11},\,\mathbf{p}_{1}+\mathbf{q}}^{+}\,{\large e}_{%
\mathbf{\Gamma }_{11},\,\mathbf{p}_{2}-\mathbf{q}}^{+}\left\{ \sum\limits_{%
\mathbf{s}_{1},\mathbf{s}_{2}}{\large h}_{\mathbf{\Gamma }_{8}\mathbf{,\,s}%
_{1}}^{+}{\large h}_{\mathbf{\Gamma }_{8}\mathbf{,\,s}_{1}}{\large h}_{%
\mathbf{\Gamma }_{8}\mathbf{,\,s}_{2}}^{+}{\large h}_{\mathbf{\Gamma }_{8}%
\mathbf{,\,s}_{2}}\right.  \notag \\
&&+\left. \sum\limits_{\mathbf{s}_{1}^{\prime },\mathbf{s}_{2}^{\prime }}%
{\large h}_{\mathbf{\Gamma }_{7}\mathbf{,\,s}_{1}^{\prime }}^{+}{\large h}_{%
\mathbf{\Gamma }_{7}\mathbf{,\,s}_{1}^{\prime }}{\large h}_{\mathbf{\Gamma }%
_{7}\mathbf{,\,s}_{2}^{\prime }}^{+}{\large h}_{\mathbf{\Gamma }_{7}\mathbf{%
,\,s}_{2}^{\prime }}+2\sum\limits_{\mathbf{s},\mathbf{s}^{\prime }}{\large h}%
_{\mathbf{\Gamma }_{8}\mathbf{,\,s}}^{+}{\large h}_{\mathbf{\Gamma }_{8}%
\mathbf{,\,s}}{\large h}_{\mathbf{\Gamma }_{7}\mathbf{,\,s}^{\prime }}^{+}%
{\large h}_{\mathbf{\Gamma }_{7}\mathbf{,\,s}^{\prime }}\right\} {\large e}_{%
\mathbf{\Gamma }_{11},\,\mathbf{p}_{2}\,}{\large e}_{\mathbf{\Gamma }_{11},\,%
\mathbf{p}_{1}}.
\end{eqnarray}%
To perform the electron-hole pairing, we first reorder the fermionic
operators taking into account the anticommutation rules. Then we apply Eqs.
(8) in the main text to products of creation operators, and their hermitic
conjugates to products of annihilation ones. For the first operator product
in (1) it looks as follows, 
\begin{eqnarray}
&&\sum\limits_{\mathbf{q},\,\mathbf{p}_{1},\,\mathbf{p}_{2},\mathbf{s}_{1},%
\mathbf{s}_{2}}V_{q}\,\underbrace{{\large e}_{\mathbf{\Gamma }_{11},\,%
\mathbf{p}_{1}+\mathbf{q}}^{+}\,\underbrace{{\large h}_{\mathbf{\Gamma }_{8}%
\mathbf{,\,s}_{1}}^{+}{\large e}_{\mathbf{\Gamma }_{11},\,\mathbf{p}_{2}-%
\mathbf{q}}^{+}}{\large h}_{\mathbf{\Gamma }_{8}\mathbf{,\,s}_{2}}^{+}}\,%
\overbrace{{\large h}_{\mathbf{\Gamma }_{8}\mathbf{,\,s}_{2}}{\large e}_{%
\mathbf{\Gamma }_{11},\,\mathbf{p}_{2}\,}}\overbrace{{\large h}_{\mathbf{%
\Gamma }_{8}\mathbf{,\,s}_{1}}{\large e}_{\mathbf{\Gamma }_{11},\,\mathbf{p}%
_{1}}}  \notag \\
&=&\frac{1}{S^{2}}\sum\limits_{\mathbf{q},\,\mathbf{p}_{1},\,\mathbf{p}_{2},%
\mathbf{s}_{1},\mathbf{s}_{2}}\left\{ \digamma \left( \alpha \mathbf{s}%
_{1}-\beta \mathbf{p}_{1}-\beta \mathbf{q}\right) \digamma \left( \alpha 
\mathbf{s}_{2}-\beta \mathbf{p}_{2}+\beta \mathbf{q}\right) A_{\Gamma _{2},\,%
\mathbf{p}_{1}+\mathbf{s}_{1}+\mathbf{q}}^{+}A_{\Gamma _{2},\,\mathbf{p}_{2}+%
\mathbf{s}_{2}-\mathbf{q}}^{+}\right.  \notag \\
&&\left. +\,\digamma \left( \alpha \mathbf{s}_{2}-\beta \mathbf{p}_{1}-\beta 
\mathbf{q}\right) \digamma \left( \alpha \mathbf{s}_{1}-\beta \mathbf{p}%
_{2}+\beta \mathbf{q}\right) A_{\Gamma _{2},\,\mathbf{p}_{1}+\mathbf{s}_{2}+%
\mathbf{q}}^{+}\left( -A_{\Gamma _{2},\,\mathbf{p}_{2}+\mathbf{s}_{1}-%
\mathbf{q}}^{+}\right) \right\}  \notag \\
&&\times \digamma \left( \alpha \mathbf{s}_{2}-\beta \mathbf{p}_{2}\right)
^{\ast }\digamma \left( \alpha \mathbf{s}_{1}-\beta \mathbf{p}_{1}\right)
^{\ast }A_{\Gamma _{2},\,\mathbf{p}_{2}+\mathbf{s}_{2}}A_{\Gamma _{2},\,\,%
\mathbf{p}_{1}+\mathbf{s}_{1}}  \notag \\
&\equiv &\sum\limits_{\mathbf{q},\,\mathbf{p}_{1},\,\mathbf{p}_{2},\mathbf{s}%
_{1},\mathbf{s}_{2}}V_{q}\underbrace{{\large e}_{\mathbf{\Gamma }_{11},\,%
\mathbf{p}_{1}+\mathbf{q}}^{+}{\large h}_{\mathbf{\Gamma }_{8}\mathbf{,\,s}%
_{1}}^{+}}\underbrace{{\large e}_{\mathbf{\Gamma }_{11},\,\mathbf{p}_{2}-%
\mathbf{q}}^{+}{\large h}_{\mathbf{\Gamma }_{8}\mathbf{,\,s}_{2}}^{+}}\,%
\overbrace{{\large h}_{\mathbf{\Gamma }_{8}\mathbf{,\,s}_{2}}\overbrace{%
{\large e_{\mathbf{\Gamma }_{11},\,\mathbf{p}_{2}\,}h}_{\mathbf{\Gamma }_{8}%
\mathbf{,\,s}_{1}}}{\large e}_{\mathbf{\Gamma }_{11},\,\mathbf{p}_{1}}}
\end{eqnarray}

\begin{figure}[tbp] 
  \centering
  \includegraphics[width=6.9in,height=2.81in,keepaspectratio]{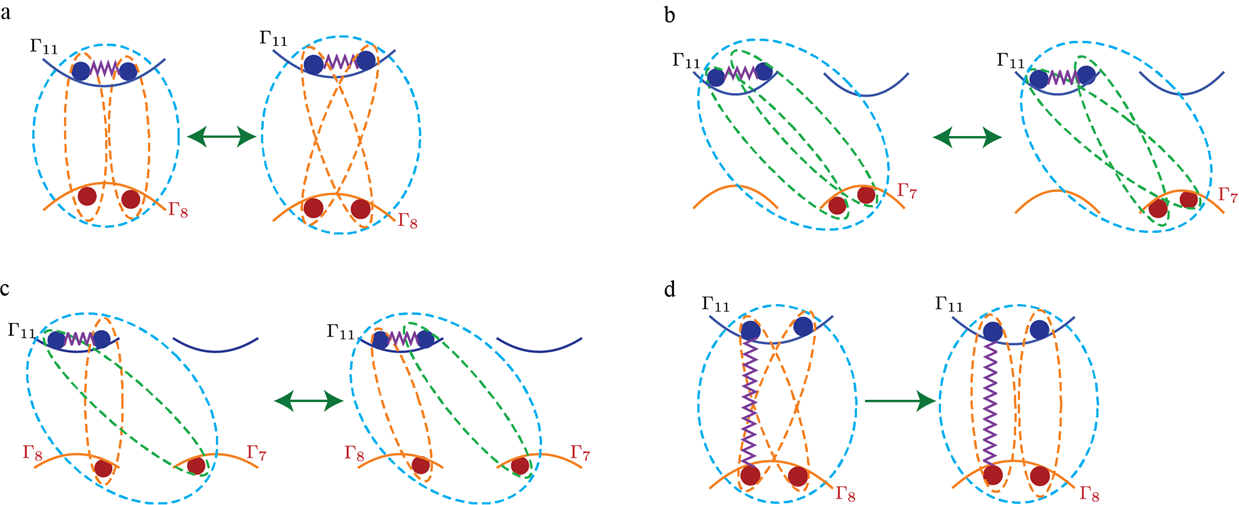}
\caption{Sketch of the direct and exchange exciton-exciton interaction
mediated by the intravalley interaction between constituent
carriers-fermions of two excitons. Light blue dashed ovals confine
correlated electron-hole pairs, the zigzag line depicts the carrier-carrier
interaction, and the rest are the same as in Fig. 1 of the main text. (a),
(d) Intravalley interaction between the $\Gamma _{2}$ excitons mediated
respectively by the intravalley electron-electron and electron-hole
interaction, and (b), (c) Interaction of the dark $\Gamma _{6}$ exciton with
an identical counterpart and a bright $\Gamma _{2}$ exciton, respectively,
mediated by the intravalley electron-electron interaction.}
\label{fig:FS1}
\end{figure}

Beginning with the annihilation operators, the pairing shown by the
horizontal curly braces over operators of pairs gives two $\Gamma _{2}$
excitons "before the interaction". If four creation operators are paired in
the same way as the annihilation ones, we have after the interaction the
same excitons, so two-pair correlations result in a direct exciton-exciton
interaction (see Fig. 1a, the left panel). If not, as shown by horizontal
braces below creation operators, the two interacting excitons exchange with
each other one of their constituents resulting in an exchange
exciton-exciton interaction (see Fig. 1a, the right panel). The product of
the exchange interaction is again two $\Gamma _{2}$ excitons. The difference
is in the functions describing the two parts of the mutual interaction of $%
\Gamma _{2}$ excitons, as one might observe comparing the two terms on the
right-hand side of Eq. (2). At least, one can see from Fig. 1a that the
momentum transfer $\mathbf{Q}$ in the direct exciton-exciton interaction is
just vector $\mathbf{q}$ transferred between two interacting electrons,
whereas in the exchange interaction, $\mathbf{Q}$ is the sum of $\mathbf{q}$
and the difference between momenta of these two electrons, $\mathbf{p}_{1}-%
\mathbf{p}_{2}$. There is another way of the pairing of two electrons and
two holes into two excitons shown by the last line in (2) that gives rise to
the reverse interaction process. It is displayed schematically in the same
Fig. 1a with the direct interaction in the right panel and the exchange one
in the left panel. Because of the indistinguishability of two electrons and
two holes, the mutually reverse interaction processes turn out to be the
same.

In a like manner, performing the pairing with the second operator product in
(1), one gets the direct and exchange interaction between two $\Gamma _{6}$
excitons, while with the third product -- between the $\Gamma _{2}$ exciton
with $\Gamma _{6}$ one. Both interaction processes are reversible, as shown
by two-side arrows in Fig. 1b and 1c. They involve carriers from different
valleys, in contrast to the intravalley processes depicted in Fig. 1a. We
see in the latter case an expected outcome, when two-pair correlations
within the $K$ valley give rise to the mutual interaction among excitons in
this valley. Here the correlations mediated by the electron-electron
interaction are shown, but in the same way they are also mediated by the
intravalley hole-hole and electron-hole interactions. To see the last case,
we repeat the procedures (1) -- (2) for the $K$ valley electron-hole
interaction term of $\mathcal{\mathcal{H}}_{eh}$. Two-pair correlations
induced by this term are presented by the following operator products, 
\begin{eqnarray}
&&\sum\limits_{\mathbf{q},\,\mathbf{p}_{1},\,\mathbf{p}_{2}}{\large e}_{%
\mathbf{\Gamma }_{11},\,\mathbf{p}_{1}+\mathbf{q}}^{+}{\large h}_{\mathbf{%
\Gamma }_{8}\mathbf{,\,p}_{2}-\mathbf{q}}^{+}\left\{ \sum\limits_{\mathbf{s}%
_{1},\mathbf{s}_{2}}{\large e}_{\mathbf{\Gamma }_{11}\mathbf{,s}_{1}}^{+}%
{\large e}_{\mathbf{\Gamma }_{11}\mathbf{,\,s}_{1}}{\large h}_{\mathbf{%
\Gamma }_{8}\mathbf{,\,s}_{2}}^{+}{\large h}_{\mathbf{\Gamma }_{8}\mathbf{%
,\,s}_{2}}+\sum\limits_{\mathbf{s}_{1}^{\prime },\mathbf{s}_{2}^{\prime }}%
{\large e}_{\mathbf{\Gamma }_{12}\mathbf{,\,s}_{1}^{\prime }}^{+}{\large e}_{%
\mathbf{\Gamma }_{12}\mathbf{,\,s}_{1}^{\prime }}{\large h}_{\mathbf{\Gamma }%
_{7}\mathbf{,\,s}_{2}^{\prime }}^{+}{\large h}_{\mathbf{\Gamma }_{7}\mathbf{%
,\,s}_{2}^{\prime }}\right.   \notag \\
&&+\left. \sum\limits_{\mathbf{s},\mathbf{s}^{\prime }}\left[ {\large e}_{%
\mathbf{\Gamma }_{12}\mathbf{,\,s}^{\prime }}^{+}{\large e}_{\mathbf{\Gamma }%
_{12}\mathbf{,\,s}^{\prime }}{\large h}_{\mathbf{\Gamma }_{8}\mathbf{,\,s}%
}^{+}{\large h}_{\mathbf{\Gamma }_{8}\mathbf{,\,s}}+{\large e}_{\mathbf{%
\Gamma }_{11}\mathbf{,s}}^{+}{\large e}_{\mathbf{\Gamma }_{11}\mathbf{,\,s}}%
{\large h}_{\mathbf{\Gamma }_{7}\mathbf{,\,s}^{\prime }}^{+}{\large h}_{%
\mathbf{\Gamma }_{7}\mathbf{,\,s}^{\prime }}\right] \right\} {\large h}_{%
\mathbf{\Gamma }_{8},\,\mathbf{p}_{2}\,}{\large e}_{\mathbf{\Gamma }_{11},\,%
\mathbf{p}_{1}}.
\end{eqnarray}%
Performing the pairing with operators of the first product, we get the
direct and exchange interaction between two excitons at the $K$ valley,
pictured in the left and right panels of Fig. 1d, respectively. The exchange
interaction arises from the pairing shown by the horizontal curly braces as
follows 
\begin{equation}
\sum\limits_{\mathbf{q},\,\mathbf{p}_{1},\,\mathbf{p}_{2},\mathbf{s}_{1},%
\mathbf{s}_{2}}V_{q}\,\underbrace{{\large e}_{\mathbf{\Gamma }_{11},\,%
\mathbf{p}_{1}+\mathbf{q}}^{+}\,{\large h}_{\mathbf{\Gamma }_{8}\mathbf{,\,p}%
_{2}-\mathbf{q}}^{+}}\,\underbrace{{\large e}_{\mathbf{\Gamma }_{11},\,%
\mathbf{s}_{1}}^{+}{\large h}_{\mathbf{\Gamma }_{8}\mathbf{,\,s}_{2}}^{+}}\,%
\overbrace{{\large h}_{\mathbf{\Gamma }_{8}\mathbf{,\,s}_{2}}\overbrace{%
{\large e}_{\mathbf{\Gamma }_{11}\mathbf{,\,s}_{1}}{\large h}_{\mathbf{%
\Gamma }_{8}\mathbf{,\,p}_{2}}}{\large e}_{\mathbf{\Gamma }_{11},\,\mathbf{p}%
_{1}}}.
\end{equation}%
One can see that there is the only pairing way of two electrons and two
holes into two excitons wherein the interacting electron and hole belong to
different pairs. Therefore the electron-hole interaction can mediate the
interaction between excitons just in the way displayed by Fig. 1d, with the
direct part in the left panel and the exchange part in the right one. This
point concerns all exciton-exciton interactions mediated by any
electron-hole interaction term of Hamiltonian $\mathcal{\mathcal{H}}_{eh}$.
In particular, the direct and exchange interaction between the $\Gamma _{2}$
and $\Gamma _{6}$ excitons, which arises from the pairing of the last
operator product in (3), can be represented by Fig. 1c, but only read from
left to right, with the electron-hole interaction between the $K$ valley
carriers instead of the electron-electron one. In parallel, the preceding
operator product gives after pairing the interaction between the $\Gamma _{2}
$ and $\Gamma _{5}$ excitons. As to the second operator product in (3), it
is noteworthy to look at the pairing,

\begin{figure}[tbp] 
  \centering
  \includegraphics[width=6.9in,height=3.86in,keepaspectratio]{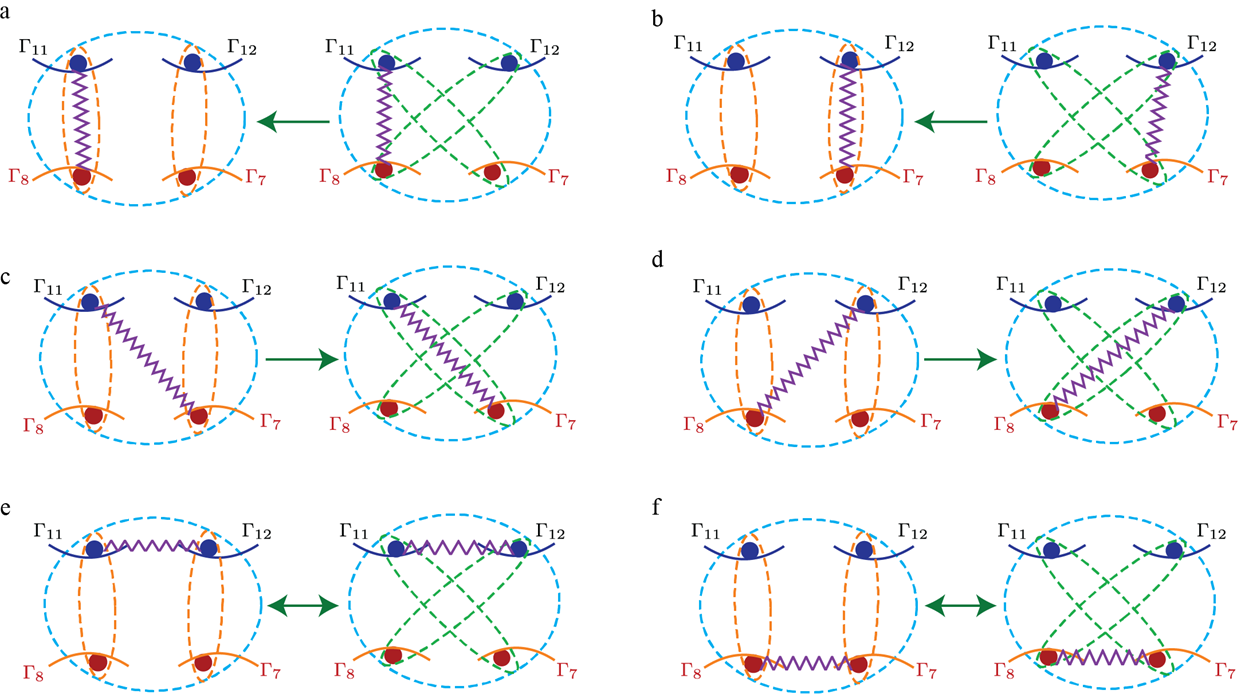}
\caption{Schematic illustration of the direct and exchange intervalley
exciton-exciton interaction mediated by: (a) and (b) -- intravalley
electron-hole interaction at the $K$ and $K^{\prime }$ valleys,
respectively, (c),(d) -- intervalley electron-hole interactions, and (e) and
(f) -- intervalley electron-electron and hole-hole interaction,
respectively. All the denotations are the same, as in Fig. 1.}
\label{fig:FS2}
\end{figure}

\begin{eqnarray}
&&\sum\limits_{\mathbf{q},\,\mathbf{p}_{1},\,\mathbf{p}_{2},\mathbf{s}%
_{1}^{\prime },\mathbf{s}_{2}^{\prime }}V_{q}\left\{ \underbrace{{\large e}_{%
\mathbf{\Gamma }_{11},\,\mathbf{p}_{1}+\mathbf{q}}^{+}\underbrace{{\large h}%
_{\mathbf{\Gamma }_{8}\mathbf{,\,p}_{2}-\mathbf{q}}^{+}{\large e}_{\mathbf{%
\Gamma }_{12}\mathbf{,\,s}_{1}^{\prime }}^{+}}{\large h}_{\mathbf{\Gamma }%
_{7}\mathbf{,\,s}_{2}^{\prime }}^{+}}\right.   \notag \\
&&-\left. \underbrace{{\large e}_{\mathbf{\Gamma }_{11},\,\mathbf{p}_{1}+%
\mathbf{q}}^{+}{\large h}_{\mathbf{\Gamma }_{8}\mathbf{,\,p}_{2}-\mathbf{q}%
}^{+}}\underbrace{{\large e}_{\mathbf{\Gamma }_{12}\mathbf{,\,s}_{1}^{\prime
}}^{+}{\large h}_{\mathbf{\Gamma }_{7}\mathbf{,\,s}_{2}^{\prime }}^{+}}%
\right\} \,\overbrace{{\large h}_{\mathbf{\Gamma }_{7}\mathbf{,\,s}%
_{2}^{\prime }}\overbrace{{\large e}_{\mathbf{\Gamma }_{12}\mathbf{,\,s}%
_{1}^{\prime }}{\large h}_{\mathbf{\Gamma }_{8}\mathbf{,\,p}_{2}}}{\large e}%
_{\mathbf{\Gamma }_{11},\,\mathbf{p}_{1}}}  \notag \\
&=&\frac{1}{S^{2}}\sum\limits_{\mathbf{q},\,\mathbf{p}_{1},\,\mathbf{p}_{2},%
\mathbf{s}_{1}^{\prime },\mathbf{s}_{2}^{\prime }}V_{q}\left\{ \digamma
\left( \alpha \mathbf{s}_{2}^{\prime }-\beta \mathbf{p}_{1}-\beta \mathbf{q}%
\right) \digamma \left( \alpha \mathbf{p}_{2}-\alpha \mathbf{q}-\beta 
\mathbf{s}_{1}^{\prime }\right) A_{\Gamma _{6},\,\mathbf{p}_{1}\mathbf{+s}%
_{2}^{\prime }+\mathbf{q}}^{+}\,\left( -A_{\Gamma _{5},\,\mathbf{p}_{2}+%
\mathbf{s}_{1}^{\prime }-\mathbf{q}}^{+}\right) \right.   \notag \\
&&\left. -\,\digamma \left( \alpha \mathbf{p}_{2}-\beta \mathbf{p}_{1}-%
\mathbf{q}\right) \digamma (\alpha \mathbf{s}_{2}^{\prime }-\beta \mathbf{s}%
_{1}^{\prime })A_{\Gamma _{2},\,\mathbf{p}_{1}+\mathbf{p}_{2}}^{+}A_{\Gamma
_{3},\,\mathbf{s}_{2}^{\prime }+\mathbf{s}_{1}^{\prime }}^{+}\right\}  
\notag \\
&&\times \digamma (\alpha \mathbf{s}_{2}^{\prime }-\beta \mathbf{p}%
_{1})^{\ast }\digamma (\alpha \mathbf{p}_{2}-\beta \mathbf{s}_{1}^{\prime
})^{\ast }\left( -A_{\Gamma _{5},\,\mathbf{p}_{2}+\mathbf{s}_{1}^{\prime
}}\right) A_{\Gamma _{6},\,\mathbf{p}_{1}+\mathbf{s}_{2}^{\prime }},
\end{eqnarray}%
yielding the direct and exchange interaction between the dark excitons.
Coming from pairing ways shown in the first and second lines on the
left-hand side of Eq. (5), respectively, they are expressed analytically by
the corresponding lines on the right-hand side. Schematically, these direct
and exchange interactions are illustrated respectively in the right and left
panels of Fig. 2a. Thus, by mediating an exchange interaction between the
dark excitons converting them into the bright ones, an intravalley
electron-hole interaction can induce the coupling between excitons from
opposite valleys. Figure 2b shows such a coupling arising from the
intravalley electron-hole interaction at the $K^{\prime }$ valley.
Naturally, all intervalley carrier-carrier interaction terms of $\mathcal{%
\mathcal{H}}_{eh}$\ induce the intervalley coupling. As shown in Figs. 2c --
f, they all mediate the direct interaction between\ two bright excitons and
convert them into the dark ones in their exchange interaction. Figures 2e
and 2f show that the intervalley electron-electron and hole-hole
interactions can induce the reverse process as well, converting two dark
excitons back into the bright ones. The electron-hole interactions cannot
[see Figs. 2c and 2d], but they have supplementary intravalley electron-hole
counterparts that do, as seen from Figs. 2a and 2b. In short, the
intervalley interaction between identical carriers generate both direct and
exchange intervalley exciton-exciton interaction, whereas their
electron-hole counterparts -- only the direct part, and the intravalley
electron-hole interactions -- only the exchange part. As the bright excitons
are optically accessible, whereas the dark excitons are not, we can consider
the latter as intermediate states in the intervalley exciton-exciton
coupling.

Analytically, repeating procedures described by (1) -- (2) and (3) -- (5)
with every term in the interaction part of Hamiltonian $\mathcal{\mathcal{H}}%
_{eh}$, then going from the electron and hole valley momenta $\mathbf{p}_{e}$%
\ and $\mathbf{p}_{h}$ to the exciton total and relative valley momenta, we
obtain a Hamiltonian of the effective exciton-exciton interaction in the
form of Eqs. (12) -- (14) in the main text. Replacing the sums over exciton
valley momenta by those over their crystal momenta, we have the Hamiltonian
in the following form

\begin{eqnarray}
\mathcal{H}_{x-x} &=&\frac{1}{2S}\sum_{\mathbf{k}_{1},\,\mathbf{k}_{2},%
\mathbf{Q}}\left\{ \sum_{\Gamma =\Gamma _{2},\Gamma _{3}}\left[ U_{\Gamma
-\Gamma }^{d}\left( \mathbf{Q}\right) +U_{\Gamma -\Gamma }^{ex}\left( 
\mathbf{k},\mathbf{Q}\right) \right] A_{\Gamma ,\mathbf{k}_{1}+\mathbf{Q}%
}^{+}A_{\Gamma ,\mathbf{k}_{2}-\mathbf{Q}}^{+}A_{\Gamma ,\mathbf{k}%
_{2}}A_{\Gamma ,\mathbf{k}_{1}}\right.  \notag \\
&&+\left[ U_{\Gamma _{5}-\Gamma _{5}}^{d}\left( \mathbf{Q}\right) +U_{\Gamma
_{5}-\Gamma _{5}}^{ex}\left( \mathbf{k},\mathbf{Q}\right) \right] A_{\Gamma
_{5},\mathbf{k}_{1}+\mathbf{K}-\mathbf{K}^{\prime }+\mathbf{Q}}^{+}A_{\Gamma
_{5},\mathbf{k}_{2}-\mathbf{K}+\mathbf{K}^{\prime }-\mathbf{Q}}^{+}A_{\Gamma
_{5},\mathbf{k}_{2}-\mathbf{K}+\mathbf{K}^{\prime }}A_{\Gamma _{5},\mathbf{k}%
_{1}+\mathbf{K}-\mathbf{K}^{\prime }}  \notag \\
&&+\left[ U_{\Gamma _{6}-\Gamma _{6}}^{d}\left( \mathbf{Q}\right) +U_{\Gamma
_{6}-\Gamma _{6}}^{ex}\left( \mathbf{k},\mathbf{Q}\right) \right] A_{\Gamma
_{6},\mathbf{k}_{1}+\mathbf{K}^{\prime }-\mathbf{K}+\mathbf{Q}}^{+}A_{\Gamma
_{6},\mathbf{k}_{2}+\mathbf{K}-\mathbf{K}^{\prime }-\mathbf{Q}}^{+}A_{\Gamma
_{6},\mathbf{k}_{2}+\mathbf{K}-\mathbf{K}^{\prime }}A_{\Gamma _{6},\mathbf{k}%
_{1}+\mathbf{K}^{\prime }-\mathbf{K}}  \notag \\
&&+2\sum\limits_{\Gamma =\Gamma _{2},\Gamma _{3}}\left[ U_{\Gamma -\Gamma
_{5}}^{d}\left( \mathbf{Q}\right) +U_{\Gamma -\Gamma _{5}}^{ex}\left( 
\mathbf{k},\mathbf{Q}\right) \right] A_{\Gamma ,\mathbf{k}_{1}+\mathbf{Q}%
}^{+}A_{\Gamma _{5},\mathbf{k}_{2}-\mathbf{K}+\mathbf{K}^{\prime }-\mathbf{Q}%
}^{+}A_{\Gamma _{5},\mathbf{k}_{2}-\mathbf{K}+\mathbf{K}^{\prime }}A_{\Gamma
,\mathbf{k}_{1}}  \notag \\
&&+2\sum\limits_{\Gamma =\Gamma _{2},\Gamma _{3}}\left[ U_{\Gamma -\Gamma
_{6}}^{d}\left( \mathbf{Q}\right) +U_{\Gamma -\Gamma _{6}}^{ex}\left( 
\mathbf{k},\mathbf{Q}\right) \right] A_{\Gamma ,\mathbf{k}_{1}+\mathbf{Q}%
}^{+}A_{\Gamma _{6},\mathbf{k}_{2}+\mathbf{K}-\mathbf{K}^{\prime }-\mathbf{Q}%
}^{+}A_{\Gamma _{6},\mathbf{k}_{2}+\mathbf{K}-\mathbf{K}^{\prime }}A_{\Gamma
,\mathbf{k}_{1}}  \notag \\
&&+2\left[ U_{\Gamma _{2}-\Gamma _{3}}^{d}\left( \mathbf{Q}\right) A_{\Gamma
_{2},\mathbf{k}_{1}+\mathbf{Q}}^{+}\,A_{\Gamma _{3},\mathbf{k}_{2}-\mathbf{Q}%
}^{+}\,A_{\Gamma _{3},\mathbf{k}_{2}}\,A_{\Gamma _{2},\mathbf{k}_{1}}\right.
\notag \\
&&\left. +U_{\Gamma _{5}-\Gamma _{6}}^{d}\left( \mathbf{Q}\right) A_{\Gamma
_{5},\mathbf{k}_{1}\mathbf{+K}-\mathbf{K}^{\prime }+\mathbf{Q}%
}^{+}\,A_{\Gamma _{6},\mathbf{k}_{2}\mathbf{+K}^{\prime }-\mathbf{K}-\mathbf{%
Q}}^{+}A_{\Gamma _{6},\mathbf{k}_{2}\mathbf{+K}^{\prime }-\mathbf{K}%
}\,A_{\Gamma _{5},\mathbf{k}_{1}\mathbf{+K}-\mathbf{K}^{\prime }}\right] 
\notag \\
&&-2\left[ U_{\Gamma _{2}-\Gamma _{3}}^{ex}\left( \mathbf{k},\mathbf{Q}%
\right) A_{\Gamma _{2},\mathbf{k}_{1}+\mathbf{Q}}^{+}\,A_{\Gamma _{3},%
\mathbf{k}_{2}-\mathbf{Q}}^{+}\,A_{\Gamma _{6},\mathbf{k}_{2}\mathbf{+K}%
^{\prime }-\mathbf{K}}\,A_{\Gamma _{5},\mathbf{k}_{1}\mathbf{+K}-\mathbf{K}%
^{\prime }}\right.  \notag \\
&&\left. \left. +U_{\Gamma _{5}-\Gamma _{6}}^{ex}\left( \mathbf{k},\mathbf{Q}%
\right) A_{\Gamma _{5},\mathbf{k}_{1}\mathbf{+K}-\mathbf{K}^{\prime }+%
\mathbf{Q}}^{+}\,A_{\Gamma _{6},\mathbf{k}_{2}\mathbf{+K}^{\prime }-\mathbf{K%
}-\mathbf{Q}}^{+}\,A_{\Gamma _{3},\mathbf{k}_{2}}\,A_{\Gamma _{2},\mathbf{k}%
_{1}}\right] \right\} ,
\end{eqnarray}%
where $\mathbf{k\equiv k}_{1}-\mathbf{k}_{2}$. In the same way, we obtain
the direct and exchange interaction energies for different symmetry
combinations in the form of sums over crystal momenta. In particular, the
exchange interaction energy between $\Gamma _{2}$ excitons reads

\begin{eqnarray}
U_{\Gamma _{2}-\Gamma _{2}}^{ex}\left( \mathbf{k},\mathbf{Q}\right) &=&-%
\frac{1}{S^{2}}\sum_{\mathbf{q}_{1},\,\mathbf{q}_{2}}V_{q_{1}}\left\{
\digamma \left( \mathbf{q}_{2}+\alpha \mathbf{Q+K}\right) \digamma \left( 
\mathbf{q}_{2}-\mathbf{q}_{1}+\beta \mathbf{k}+\beta \mathbf{Q}+\mathbf{K}%
\right) \,\digamma \left( \mathbf{q}_{2}-\mathbf{q}_{1}+\beta \mathbf{k}+%
\mathbf{Q}+\mathbf{K}\right) ^{\ast }\right.  \notag \\
&&+\digamma \left( \mathbf{q}_{2}-\beta \mathbf{Q}+\mathbf{K}\right)
\digamma \left( \mathbf{q}_{2}-\mathbf{q}_{1}-\alpha \mathbf{k}-\alpha 
\mathbf{Q}+\mathbf{K}\right) \digamma \left( \mathbf{q}_{2}-\mathbf{q}%
_{1}-\alpha \mathbf{k}-\mathbf{Q}+\mathbf{K}\right) ^{\ast }  \notag \\
&&-2\digamma \left( \mathbf{q}_{2}-\mathbf{q}_{1}+\alpha \mathbf{Q+K}\right)
\digamma \left( \mathbf{q}_{2}-\mathbf{q}_{1}+\beta \mathbf{k}+\beta \mathbf{%
Q}+\mathbf{K}\right)  \notag \\
&&\left. \times \,\digamma \left( \mathbf{q}_{2}-\mathbf{q}_{1}+\beta 
\mathbf{k}+\mathbf{Q}+\mathbf{K}\right) ^{\ast }\right\} \digamma \left( 
\mathbf{q}_{2}+\mathbf{K}\right) ^{\ast }
\end{eqnarray}%
Replacing vector $\mathbf{K}$ by $\mathbf{K}^{\prime }$, $\alpha \mathbf{K}%
+\beta \mathbf{K}^{\prime }$, and $\alpha \mathbf{K}^{\prime }+\beta \mathbf{%
K}$, we get from here energies $U_{\Gamma _{3}-\Gamma _{3}}^{ex}\left( 
\mathbf{k},\mathbf{Q}\right) $, $U_{\Gamma _{5}-\Gamma _{5}}^{ex}\left( 
\mathbf{k},\mathbf{Q}\right) $, and $U_{\Gamma _{6}-\Gamma _{6}}^{ex}\left( 
\mathbf{k},\mathbf{Q}\right) $ of the exchange interaction between identical 
$\Gamma _{3}$, $\Gamma _{5}$, and $\Gamma _{6}$ excitons, respectively.

For the interaction between bright and dark excitons, e.g. between the $%
\Gamma _{2}$ and $\Gamma _{5}$ ones, we have

\begin{eqnarray}
U_{\Gamma _{2}-\Gamma _{5}}^{ex}\left( \mathbf{k},\mathbf{Q}\right) &=&-%
\frac{1}{S^{2}}\sum_{\mathbf{q}_{1},\,\mathbf{q}_{2}}V_{q_{1}}\left\{
\digamma \left( \mathbf{q}_{2}+\alpha \mathbf{Q+K}\right) \digamma \left( 
\mathbf{q}_{2}-\mathbf{q}_{1}+\beta \mathbf{k}+\beta \mathbf{Q}+\alpha 
\mathbf{K}+\beta \mathbf{K}^{\prime }\right) \,\right.  \notag \\
&&\times \digamma \left( \mathbf{q}_{2}-\mathbf{q}_{1}+\beta \mathbf{k}+%
\mathbf{Q}+\alpha \mathbf{K}+\beta \mathbf{K}^{\prime }\right) ^{\ast } 
\notag \\
&&+\digamma \left( \mathbf{q}_{2}-\beta \mathbf{Q}+\mathbf{K}\right)
\digamma \left( \mathbf{q}_{2}-\mathbf{q}_{1}-\alpha \mathbf{k}-\alpha 
\mathbf{Q}+\alpha \mathbf{K}+\beta \mathbf{K}^{\prime }\right)  \notag \\
&&\times \digamma \left( \mathbf{q}_{2}-\mathbf{q}_{1}-\alpha \mathbf{k}-%
\mathbf{Q}+\alpha \mathbf{K}+\beta \mathbf{K}^{\prime }\right) ^{\ast } 
\notag \\
&&-2\digamma \left( \mathbf{q}_{2}-\mathbf{q}_{1}+\alpha \mathbf{Q+K}\right)
\digamma \left( \mathbf{q}_{2}-\mathbf{q}_{1}+\beta \mathbf{k}+\beta \mathbf{%
Q}+\alpha \mathbf{K}+\beta \mathbf{K}^{\prime }\right)  \notag \\
&&\left. \times \,\digamma \left( \mathbf{q}_{2}-\mathbf{q}_{1}+\beta 
\mathbf{k}+\mathbf{Q}+\alpha \mathbf{K}+\beta \mathbf{K}^{\prime }\right)
^{\ast }\right\} \digamma \left( \mathbf{q}_{2}+\mathbf{K}\right) ^{\ast }
\end{eqnarray}

Lastly, for the exchange interaction energy between bright $\Gamma _{2}$ and 
$\Gamma _{3}$ excitons, 
\begin{eqnarray}
U_{\Gamma _{2}-\Gamma _{3}}^{ex}\left( \mathbf{k},\mathbf{Q}\right) &=&-%
\frac{1}{S^{2}}\sum_{\mathbf{q}_{1},\,\mathbf{q}_{2}}V_{q_{1}}\left\{
\digamma \left( \mathbf{q}_{2}-\mathbf{q}_{1}+\beta \mathbf{k}+\beta \mathbf{%
Q}+\mathbf{K}\right) \digamma \left( \mathbf{q}_{2}+\alpha \mathbf{Q+K}%
^{\prime }\right) \,\right.  \notag \\
&&\times \digamma \left( \mathbf{q}_{2}-\mathbf{q}_{1}+\beta \mathbf{k}+%
\mathbf{Q}+\alpha \mathbf{K}+\beta \mathbf{K}^{\prime }\right) ^{\ast } 
\notag \\
&&+\digamma \left( \mathbf{q}_{2}-\mathbf{q}_{1}-\alpha \mathbf{k}-\alpha 
\mathbf{Q}+\mathbf{K}\right) \digamma \left( \mathbf{q}_{2}-\beta \mathbf{Q}+%
\mathbf{K}^{\prime }\right)  \notag \\
&&\times \digamma \left( \mathbf{q}_{2}-\mathbf{q}_{1}-\alpha \mathbf{k}-%
\mathbf{Q}+\alpha \mathbf{K}+\beta \mathbf{K}^{\prime }\right) ^{\ast } 
\notag \\
&&-2\digamma \left( \mathbf{q}_{2}-\mathbf{q}_{1}+\beta \mathbf{k}+\beta 
\mathbf{Q}+\mathbf{K}\right) \digamma \left( \mathbf{q}_{2}-\mathbf{q}%
_{1}+\alpha \mathbf{Q+K}^{\prime }\right)  \notag \\
&&\left. \times \,\digamma \left( \mathbf{q}_{2}-\mathbf{q}_{1}+\beta 
\mathbf{k}+\mathbf{Q}+\alpha \mathbf{K}+\beta \mathbf{K}^{\prime }\right)
^{\ast }\right\} \digamma \left( \mathbf{q}_{2}+\alpha \mathbf{K}^{\prime
}+\beta \mathbf{K}\right) ^{\ast }
\end{eqnarray}%
and that of the reverse process,%
\begin{eqnarray}
U_{\Gamma _{5}-\Gamma _{6}}^{ex}\left( \mathbf{k},\mathbf{Q}\right) &=&-%
\frac{1}{S^{2}}\sum_{\mathbf{q}_{1},\,\mathbf{q}_{2}}V_{q_{1}}\left\{
\digamma \left( \mathbf{q}_{2}-\mathbf{q}_{1}+\beta \mathbf{k}+\beta \mathbf{%
Q}+\alpha \mathbf{K}+\beta \mathbf{K}^{\prime }\right) \digamma \left( 
\mathbf{q}_{2}+\alpha \mathbf{Q}+\alpha \mathbf{K}^{\prime }+\beta \mathbf{K}%
\right) \,\right.  \notag \\
&&\times \digamma \left( \mathbf{q}_{2}-\mathbf{q}_{1}+\beta \mathbf{k}+%
\mathbf{Q}+\mathbf{K}\right) ^{\ast }  \notag \\
&&+\digamma \left( \mathbf{q}_{2}-\mathbf{q}_{1}-\alpha \mathbf{k}-\alpha 
\mathbf{Q}+\alpha \mathbf{K}+\beta \mathbf{K}^{\prime }\right) \digamma
\left( \mathbf{q}_{2}-\beta \mathbf{Q}+\alpha \mathbf{K}^{\prime }+\beta 
\mathbf{K}\right)  \notag \\
&&\times \digamma \left( \mathbf{q}_{2}-\mathbf{q}_{1}-\alpha \mathbf{k}-%
\mathbf{Q}+\mathbf{K}\right) ^{\ast }  \notag \\
&&-2\digamma \left( \mathbf{q}_{2}-\mathbf{q}_{1}+\beta \mathbf{k}+\beta 
\mathbf{Q}+\alpha \mathbf{K}+\beta \mathbf{K}^{\prime }\right) \digamma
\left( \mathbf{q}_{2}-\mathbf{q}_{1}+\alpha \mathbf{Q}+\alpha \mathbf{K}%
^{\prime }+\beta \mathbf{K}\right)  \notag \\
&&\left. \times \,\digamma \left( \mathbf{q}_{2}-\mathbf{q}_{1}+\beta 
\mathbf{k}+\mathbf{Q}+\mathbf{K}\right) ^{\ast }\right\} \digamma \left( 
\mathbf{q}_{2}+\mathbf{K}^{\prime }\right) ^{\ast }.
\end{eqnarray}